\documentclass[12pt]{article}
\pdfoutput=1
\usepackage{amsfonts,hyperref}
\usepackage{color}
\usepackage{appendix}
\usepackage{amsmath}
\usepackage{graphicx}
\usepackage[latin1]{inputenc}
\usepackage[french,english]{babel}
\usepackage{geometry}
\usepackage{etoolbox}
\usepackage{fancyhdr}
\patchcmd{\thebibliography}{\section*}{\section}{}{}

\addto\captionsfrench{}
\addto\captionsenglish{}

\newcommand\encadremath[1]{\vbox{\hrule\hbox{\vrule\kern8pt
\vbox{\kern8pt \hbox{$\displaystyle #1$}\kern8pt}
\kern8pt\vrule}\hrule}}
\def\enca#1{\vbox{\hrule\hbox{
\vrule\kern8pt\vbox{\kern8pt \hbox{$\displaystyle #1$}
\kern8pt} \kern8pt\vrule}\hrule}}

\newcommand\figureframex[3]{
\begin{figure}[bth]
\hrule\hbox{\vrule\kern8pt
\vbox{\kern8pt \vbox{
\begin{center}
{\mbox{\epsfxsize=#1.truecm\epsfbox{#2}}}
\end{center}
\caption{#3}
}\kern8pt}
\kern8pt\vrule}\hrule
\end{figure}
}
\makeatletter
\@addtoreset{equation}{section}
\makeatother
\newtheorem{theorem}{Theorem}[section]

\newtheorem{remark}{Remark}[section]
\newtheorem{proposition}{Proposition}[section]
\newtheorem{lemma}{Lemma}[section]
\newtheorem{corollary}{Corollary}[section]
\newtheorem{definition}{Definition}[section]

\def\br{\begin{remark}\rm\small}
\def\er{\end{remark}}
\def\bt{\begin{theorem}}
\def\et{\end{theorem}}
\def\bd{\begin{definition}}
\def\ed{\end{definition}}
\def\bp{\begin{proposition}}
\def\ep{\end{proposition}}
\def\bl{\begin{lemma}}
\def\el{\end{lemma}}
\def\bc{\begin{corollary}}
\def\ec{\end{corollary}}
\def\beaq{\begin{eqnarray}}
\def\eeaq{\end{eqnarray}}

\newcommand{\eq}[1]{eq.~(\ref{#1})}

\newcommand{\beq}{\begin{equation}}
\newcommand{\eeq}{\end{equation}}
\newcommand{\bea}{\begin{eqnarray}}
\newcommand{\eea}{\end{eqnarray}}

 \newcommand{\Tr}{{\,\rm Tr}\:}

\newcommand{\td}[1]{{\tilde{#1}}}

\newcommand{\e}{{\,\rm e}\,}
\newcommand{\ee}[1]{{{\rm e}^{#1}}}

\newcommand{\Pint}{{\int\kern -1.em -\kern-.25em}}

\newcommand\Res{\mathop{{\rm Res}}}

\begin{document}

\sloppy


\pagestyle{empty}
\hfill{IPHT: t13/251~}
\linebreak
\hspace*{0pt}
\hfill{CRM-3329-2013}
\vspace{10pt}
\begin{center}
{\large \bf {The sine-law gap probability, Painlevé $5$, and asymptotic expansion by the topological recursion.}}
\end{center}
\vspace{5pt}
\begin{center}
\textbf{O. Marchal$^\dagger$, B. Eynard$^\ddagger$, M. Bergère$^\star$}
\end{center}
\vspace{20pt}

$^\dagger$ \textit{Université de Lyon, CNRS UMR 5208, Université Jean Monnet, Institut Camille Jordan, France}
\footnote{olivier.marchal@univ-st-etienne.fr}

$^\ddagger$ \textit{Institut de physique théorique, CEA Saclay, France et Centre de Recherche Mathématiques, Montréal.}
\footnote{bertrand.eynard@cea.fr}

$^\star$ \textit{Institut de physique théorique, CEA Saclay, France.}
\footnote{michel.bergere@cea.fr}

\vspace{30pt}

{\bf Abstract}:
The goal of this article is to rederive the connection between the Painlevé $5$ integrable system and the universal eigenvalues correlation functions of double-scaled hermitian matrix models, through the topological recursion method. More specifically we prove, \textbf{to all orders}, that the WKB asymptotic expansions of the $\tau$-function as well as of determinantal formulas arising from the Painlevé $5$ Lax pair are identical to the large $N$ double scaling asymptotic expansions of the partition function and correlation functions of any hermitian matrix model around a regular point in the bulk. In other words, we rederive the ``sine-law" universal bulk asymptotic of large random matrices and provide an alternative perturbative proof of universality in the bulk with only algebraic methods. Eventually we exhibit the first orders of the series expansion up to $O(N^{-5})$.

\tableofcontents

\vspace{30pt}
\pagestyle{plain}
\setcounter{page}{1}


\section{Introduction}

We consider the standard hermitian matrix integrals given by the partition function:
\beq Z_n=\int\dots\int d\lambda |\Delta(\lambda)|^2e^{-\frac{N}{T}\underset{i=1}{\overset{n}{\sum}} V(\lambda_i)}\eeq
It is well known in the literature \cite{MehtaBook,JMI} that the correlation functions in the bulk part of the spectrum are closely related to Fredholm determinants given by the sine kernel. Moreover, it is also well known that these Fredholm determinants can be rewritten with the use of Painlevé transcendents. For example, the probability that no eigenvalues lie in an interval $I=[0,t]$ is connected to a solution of the $\sigma$-form of the Painlevé $5$ equation. In this article, we plan to apply the topological recursion method, successfully applied for the Airy kernel and Painlevé $2$ case in \cite{MC, BorEyn1,BorEyn2,BEMN}, to give a new proof of this relation. The general strategy is the following:
\begin{enumerate}
\item On the matrix model side, we define a partition function as the probability that no eigenvalues lie in a given interval inside the bulk, and take its scaling limit for which we write explicitly the spectral curve. Then, the topological recursion developed in \cite{EO} gives a recursive way to write all correlation functions and the $\frac{1}{N}$ expansion of the partition function known as the symplectic invariants $F^{(g)}$.
\item On the integrable system side, since the works Jimbo, Miwa and Sato \cite{JMI,JMII} we know the Lax pair corresponding to the Painlevé $5$ equation. Moreover, from the works of \cite{BE}, we know how to define correlation functions associated to any $2\times 2$ Lax pair by determinantal formulas. These determinantal formulas obey loop equations, which imply, under some appropriate analytical assumptions, that their asymptotic expansions can be computed by the topological recursion of \cite{EO}, applied to the Lax pair's spectral curve.
\item In both formalism (Lax pair and matrix model), we explain why quantities involved ($\ln \tau$ or partition function) have a series expansion in the parameter $\hbar$ (connected to $\frac{1}{N}$ on the matrix model side). On the integrable system side, this was already known for the case of vanishing monodromy which is the case arising in this paper. On the matrix model side, after scaling, we obtain a genus $0$ curve for which the standard theory of the topological recursion proves that the correlation functions have a series expansion in most cases. 
\item We check that the spectral curve of Painlevé $5$ matches the one obtained from the scaling limit of the gap probability. Then we validate the analytical assumptions needed for matching the correlation functions built on the Lax pair to the one generated by the topological recursion on the matrix model side (Cf. \ref{MainTheo}). We stress here that this verification is essential since even if the correlation functions built by the topological recursion on the spectral curve and the correlation functions built on the Lax pairs are known to satisfy the loop equations, these loop equations have many solutions and therefore one needs to validate several conditions in order to ensure that these two sets of solutions are identical. At the technical level, verifying the three analytical conditions (presented in appendix \ref{AppendixB}, \ref{AppendixParity} and \ref{AppendixthWnleading}) is far from being trivial and each of them implies the use of sophisticated arguments about the underlying structure of the integrable system. 
\item In the end we are able to conclude that correlation functions computed on both sides (matrix model and Painlevé $5$) must match to all orders. In other words, after identifying the spectral curves and the pole structure, we get uniqueness of the solutions of the loop equations (satisfied in both cases) and thus we conclude that quantities on both sides are identical to all orders. In this way, we reconnect the Painlevé $5$ solution to the scaling limit of matrix integral in a more direct way. It also complements similar results obtained recently for the other known integrable cases: eigenvalues statistics at the end of the distribution, merging of two intervals, etc. that were studied with a similar method in \cite{BorEyn1,BleherEynard,MC}. 
\end{enumerate}

\section{\label{section1}Reminder about universality in random matrices}

It has been known for a while that the gap probability, i.e. the probability of finding no eigenvalues of an hermitian random matrix in a given interval, is generically (i.e. local statistics around a regular bulk point) connected to the Painlevé $5$ integrable system. The most classical way to present the results are the one presented in \cite{MehtaBook} which we will rapidly review. Let us define $\td{\sigma}(s)$ the unique solution of the $\sigma$-form of the Painlevé $5$ equation:
\beq \label{PAINLEVE5} (s\ddot{\td{\sigma}})^2+4(s\dot{\td{\sigma}}-\td{\sigma})(s\dot{\td{\sigma}}-\td{\sigma}+(\dot{\td{\sigma}})^2)=0 \eeq
with the behavior $\td{\sigma}(s)\underset{s\to 0}{\sim}-\frac{s}{\pi}-\left(\frac{s}{\pi}\right)^2$. Then one can define a so-called $\tau$-function:
\beq \ln \td{\tau}(s)=\int_0^s \frac{\td{\sigma}(u)}{u}du\eeq
from which one can deduce the gap probability of the interval $[0,s]$ by:
\beq 
E_2(0,s)=\text{exp}\left(\int_0^{\pi s}\frac{\td{\sigma}(u)}{u}du\right)\eeq
In other words:
\beq E_2(0,s)=\td{\tau}(\pi s)\eeq
Since the variable involved in the $\tau$ function is $\pi s$ but not directly $s$ itself, it seems appropriate to perform the change of variable $s\to \frac{s}{\pi}$. Define $\hat{\sigma}(s)=\td{\sigma}(\pi s)$ then $\hat{\sigma}(s)$ is the unique solution of the equation (we still denote with a dot the derivative relatively to $s$):
\beq \label{P5bis} \frac{1}{\pi^2}(s\ddot{\hat{\sigma}})^2+4(s\dot{\hat{\sigma}}-\hat{\sigma})(s\dot{\hat{\sigma}}-\hat{\sigma}+\frac{1}{\pi^2}(\dot{\hat{\sigma}})^2)=0 \eeq
with a behavior at small $s$ given by $\hat{\sigma}(s)\underset{s\to 0}{\sim}-s-s^2$. Defining $\hat{\tau}(s)$ the corresponding $\tau$-function associated to this solution by:
\beq \ln \hat{\tau}(s)=\int_0^s \frac{\hat{\sigma}(u)}{u}du\eeq
we get that the gap probability is given by:
\beq E_2(0,s)=\hat{\tau}(s)\eeq
Hence, regarding normalization for the gap probability, we find more convenient to use the representation \eqref{P5bis} of the $\sigma$-form of the Painlevé $5$ equation for which we shall present in the next section a standard Lax pair. 

\section{Painlevé $5$ side}

In this section, we review the general formalism corresponding to the Painlevé $5$ equation. Our case will cover a generalized version of Painlevé $5$ with the addition of a natural small parameter $-\pi \hbar$ that will be matched later with the $\frac{1}{N}$ parameter arising in matrix models. Of course, one can easily recover the standard case of Painlevé $5$ usually presented in the literature by taking $\hbar=-\frac{1}{\pi}$. Note here that the factor $-\pi$ is purely conventional and is introduced here only to recover the same normalization introduced in the previous section. Our presentation here will mainly follow the one of N. Joshi, A. V. Kitaev, and P. A. Treharne in their paper ``On the Linearization of the Painlevé III-VI Equations and Reductions of the Three-Wave Resonant System" \cite{JKT}, although our goal is different from their.

\subsection{The Painlevé $5$ Lax system}
 
Jimbo-Miwa-Sato-Ueno in \cite{JMI}, \cite{JMII} give the following $2\times 2$ Lax pair:  
 
\bea \label{Lax} -\pi \hbar\frac{ \partial \Psi}{\partial \xi}(t,\xi) &=&\left( \frac{t}{2}\sigma_3 +\frac{A_0}{\xi}+\frac{A_1}{\xi-1}\right) \Psi(t,\xi)\overset{\text{def}}{=}\mathcal{D}(t,\xi)\Psi(t,\xi)\cr
 -\pi \hbar\frac{ \partial \Psi}{\partial t}(t,\xi) &=&\left( \frac{\xi}{2}\sigma_3 +\frac{1}{t}\left(A_0+A_1+\frac{\theta_\infty}{2}\sigma_3\right)\right) \Psi(t,\xi)\overset{\text{def}}{=}\mathcal{R}(t,\xi)\Psi(t,\xi)\cr
 \eea
with $\sigma_3$ the diagonal Pauli matrix and
\beq A_0=\begin{pmatrix}
z+\frac{\theta_0}{2}& -u (z+\theta_0)\\
\frac{z}{u}& -z-\frac{\theta_0}{2}\\
\end{pmatrix}
\,\, ;\,\,
A_1=\begin{pmatrix}
-z-\frac{\theta_0+\theta_\infty}{2}& uy\left(z+\frac{\theta_0-\theta_1+\theta_\infty}{2}\right)\\
-\frac{1}{uy}\left(z+\frac{\theta_0+\theta_1+\theta_\infty}{2}\right)& z+\frac{\theta_0+\theta_\infty}{2}\\
\end{pmatrix}
\eeq
where $z=z(t)$, $u=u(t)$ and $y=y(t)$ are functions of the time $t$, whereas $\theta_i$ are constant monodromy parameters:
\beq
\Tr A_0^2 = \frac{\theta_0^2}{2}
\qquad , \qquad
\Tr A_1^2 = \frac{\theta_1^2}{2}
\eeq
In other words:
\beq \mathcal{D}(t,\xi)=\left( \begin {array}{cc} \frac{t}{2}+{\frac {z+\frac{1}{2}\,\theta_{{0}}}{\xi}}-{\frac {z+\frac{1}{2}\,\theta_{{0}}+\frac{1}{2}\,\theta_{{\infty}}}{\xi-1}}&-{\frac {u \left( 
z+\theta_{{0}} \right) }{\xi}}+{\frac {uy \left( z+\frac{1}{2}\,\theta_{{0}}-\frac{1}{2}
\,\theta_{{1}}+\frac{1}{2}\,\theta_{{\infty}} \right) }{\xi-1}}\\\noalign{\medskip}{
\frac {z}{u\xi}}-{\frac {z+\frac{1}{2}\,\theta_{{0}}+\frac{1}{2}\,\theta_{{1}}+\frac{1}{2}\,
\theta_{{\infty}}}{ \left( \xi-1 \right) uy}}&-\frac{t}{2}-{\frac {z+\frac{1}{2}\,\theta_{
{0}}}{\xi}}+{\frac {z+\frac{1}{2}\,\theta_{{0}}+\frac{1}{2}\,\theta_{{\infty}}}{\xi-1}}
\end {array} \right) 
\eeq
and
\beq \mathcal{R}(t,\xi)=\left( \begin {array}{cc} \frac{\xi}{2}&-{\frac {u \left( z+\theta_{{0}} \right) }{t}}+{\frac {uy \left( z+\frac{1}{2}\,\theta_{{0}}-\frac{1}{2}\,\theta_{{1}}
+\frac{1}{2}\,\theta_{{\infty}} \right) }{t}}\\\noalign{\medskip}{\frac {z}{t u}}-{
\frac {z+\frac{1}{2}\,\theta_{{0}}+\frac{1}{2}\,\theta_{{1}}+\frac{1}{2}\,\theta_{{\infty}}}{t uy}}&
-\frac{\xi}{2}\end {array} \right) 
\eeq 

It can be proved (See for example Appendix A of \cite{JKT}) that the compatibility equation of \eqref{Lax} is
\beq
-\pi \left(\hbar\frac{\partial {\cal D}(t,\xi)}{\partial t}- \hbar\frac{\partial {\cal R}(t,\xi)}{\partial \xi}\right) = [{\cal R}(t,\xi),{\cal D}(t,\xi)]
\eeq
which gives the differential system:
\bea \label{sysyzu}-\pi \hbar t\frac{d y}{dt}&=& ty -2z(y-1)^2-(y-1)\left( \frac{\theta_0-\theta_1+\theta_\infty}{2}y -\frac{3\theta_0+\theta_1+\theta_\infty}{2}\right)\cr
-\pi \hbar t\frac{d z}{dt}&=& yz\left(z+\frac{\theta_0-\theta_1+\theta_\infty}{2}\right)-\frac{1}{y}\left(z+\theta_0\right)\left(z+\frac{\theta_0+\theta_1+\theta_\infty}{2}\right)\cr
-\pi \hbar t \frac{ d \log u}{dt}&=& -2z-\theta_0+y\left(z+\frac{\theta_0-\theta_1+\theta_\infty}{2}\right)+\frac{1}{y}\left(z+\frac{\theta_0+\theta_1+\theta_\infty}{2}\right)\cr
\eea
In order to derive the Painlevé $5$ differential equation, it is standard to introduce the function $\sigma(t)$:
\beq \label{sigma} \sigma(t)=-2z^2-z(2\theta_0+\theta_\infty+ t) +yz\left(z+ \frac{\theta_0-\theta_1+\theta_\infty}{2}\right)+\frac{\left(z+\theta_0\right)}{y}\left(z+\frac{\theta_0+\theta_1+\theta_\infty}{2}\right)\eeq
A straightforward computation shows that it satisfies the important relation:
\beq \label{sigma2}\encadremath{ \dot{\sigma}(t)=-z(t) }\eeq
We note here that there exists also another way to introduce the $\sigma(t)$ function by using the description of the Painlevé system as a $3\times 3$ matrix differential system. This approach is developed in details in \cite{JKT} and we refer the interested reader to the article and its appendices. In the perspective of the recent development of $\beta$-ensembles it may also appear that the $3\times 3$ matrix systems are more natural than the $2\times 2$ Lax pair.

Using the second equation of \eqref{sysyzu} we get:
$$\left\{ \label{bla}
    \begin{array}{ll}
        -\pi \hbar t \ddot{\sigma}+\left(\sigma+2(\dot{\sigma})^2-\left(2\theta_0+\theta_\infty+ t\right)\dot{\sigma}\right)&=\frac{2}{y}\left(\dot{\sigma}- \theta_0\right)\left(\dot{\sigma}- \frac{\theta_0+\theta_1+\theta_\infty}{2}\right) \\
        -\pi \hbar t \ddot{\sigma}-\left(\sigma+2(\dot{\sigma})^2-\left(2\theta_0+\theta_\infty+ t\right)\dot{\sigma}\right)&=-2y\dot{\sigma}\left(\dot{\sigma}- \frac{\theta_0-\theta_1+\theta_\infty}{2}\right)
    \end{array}
\right.
$$
Multiplying the last two equations together leads to the $\sigma$-form of a Painlevé $5$ (with the additional $\hbar$ parameter):

\begin{equation}
\label{P5}
   \fbox{$
   \begin{array}{rcl}
&&-\left(\pi \hbar t\ddot{\sigma}\right)^2+\left(\sigma+2(\dot{\sigma})^2-(2\theta_0+\theta_\infty+ t)\dot{\sigma}\right)^2=\cr
&&4\dot{\sigma}(\dot{\sigma}-\theta_0)\left(\dot{\sigma}- \frac{\theta_0-\theta_1+\theta_\infty}{2}\right)\left(\dot{\sigma}- \frac{\theta_0+\theta_1+\theta_\infty}{2}\right)    \end{array}
   $}
\end{equation}

We note here that this $2\times 2$ Lax pair has the advantage to be more directly useful in the correspondence of integrable systems and matrix models. However, for some computations, it is sometimes easier to use the $3\times 3$ formalism presented in \cite{JKT}.

\subsection{Vanishing monodromies}

In this article, we shall be interested in the case where all monodromies $\theta_i$ vanish:
\beq
\theta_0=\theta_1=\theta_\infty = 0.
\eeq
This requirement will simplify most of the formulas but if one wants to study other quantities than the gap probability, as for example a fixed proportion of eigenvalues, it will require the use of the total Painlevé $5$ system for the direct adaptation of our proof. This is the reason why we kept the full Painlevé $5$ system before specializing it to our case.

In that case, the Lax pair simplifies into:
\beq\label{defDtxi}
\mathcal{D}(t,\xi)=\left( \begin {array}{cc} \frac{t}{2}+{\frac {z}{\xi}}-{\frac {z}{\xi-1}}&-{\frac {u z }{\xi}}+{\frac {uy z  }{\xi-1}}\\\noalign{\medskip}{
\frac {z}{u\xi}}-{\frac {z}{ \left( \xi-1 \right) uy}}&-\frac{t}{2}-{\frac {z}{\xi}}+{\frac {z}{\xi-1}}
\end {array} \right) 
\eeq
and
\beq\label{defRtxi} \mathcal{R}(t,\xi)=\left( \begin {array}{cc} \frac{\xi}{2}&-{\frac {u z }{t}}+{\frac {uyz  }{t}}\\\noalign{\medskip}{\frac {z}{t u}}-{
\frac {z}{t uy}}&
-\frac{\xi}{2}\end {array} \right) 
\eeq 
The Chazy equation for $\sigma(t)$ reduces to
\beq \label{Chazy}
\encadremath{-( \pi\hbar t \ddot{\sigma})^2+(t\dot{\sigma}-\sigma)(t\dot{\sigma}-\sigma-4\dot{\sigma}^2)=0}
\eeq
and the connection between $\sigma(t)$ and $(y(t),z(t))$ \eqref{sigma} simplifies into:
\beq \label{reducedsigma}\sigma(t)=-2z^2-tz+yz^2+\frac{z^2}{y}\eeq

\subsection{Proper normalization to recover the standard universality formalism}

So far we have used by simplicity the Lax pair given by Jimbo-Miwa-Sato-Ueno found in \cite{JKT}. Unfortunately, even in the case without monodromy, one can see that the normalization chosen for the $\sigma(t)$ function is different from $\hat{\sigma}(s)$ presented in \eqref{P5bis} from which universality results like the gap probability and the correctly normalized $\tau$-function are simply expressed. However it is quite simple to observe that the two differential equations \eqref{P5bis} and \eqref{Chazy} are essentially the same and linked through the correspondence:
\beq \label{ChangeNormalization} s=\alpha t=\frac{i t}{2\pi} \,\,,\,\, \hat{\sigma}(s)=\sigma\left(\frac{1}{\alpha} s\right)\,\,,\,\, \hbar= -\frac{1}{\pi}\eeq
Indeed, we observe by homogeneity that $t\frac{d}{dt}\sigma(t)=s\frac{d}{ds}\hat{\sigma}(s)$ and $t^2\frac{d^2}{dt^2}\sigma(t)=s^2\frac{d^2}{ds^2}\hat{\sigma}(s)$ thus the only modified terms are the coefficient in front of $\dot{\sigma}(t)^2$ changing from $-4$ to $-4\alpha^2$ and also the coefficient in front of $t^2\left(\ddot{\sigma}(t)\right)^2$ which is also multiplied by $\alpha^2$. Taking $\alpha=\frac{i}{2\pi}$ gives the coefficient in front of $\dot{\hat{\sigma}}^2$ to be $\frac{1}{\pi^2}$ as wanted. The coefficient in front of $t^2\ddot{\sigma}(t)$ then becomes $\frac{1}{4}$ and hence multiplying by $4$ we recover the standard form given in \eqref{P5bis}.
Hence with this rescaling we find that saying that $\sigma(t)$ satisfies \eqref{Chazy} is completely equivalent to saying that $\hat{\sigma}(s)$ satisfies \eqref{P5bis}:
\beq \frac{1}{\pi^2}(s\ddot{\hat{\sigma}})^2+4(s\dot{\hat{\sigma}}-\hat{\sigma})(s\dot{\hat{\sigma}}-\hat{\sigma}+\frac{1}{\pi^2}(\dot{\hat{\sigma}})^2)=0 \eeq
We stress again that the $-\pi$ coefficient in front of $\hbar$ is only a matter of normalization that can be easily handled with simple multiplicative renormalization of time and functions in all formalism (matrix models, Lax pair and differential equation).

\subsection{Spectral curve arising in the Painlevé $5$ formalism}

With the help of the Lax Pair \eqref{Lax}, we can apply standard results between matrix models and integrable systems. Indeed it has been shown in \cite{BE} how to construct the spectral curve, correlation functions and the $\tau$-function when dealing with a $2 \times 2$ Lax pair like \eqref{Lax} coming from integrable systems. We also remark that this work have been recently generalized for any $d \times d$ Lax pair in \cite{BBEnew}. The spectral curve is defined as the limiting (when $\hbar\to 0$) characteristic polynomial of the $-\frac{1}{\pi}\mathcal{D}(t,\xi)$ matrix (the factor $-\frac{1}{\pi}$ is present since the spectral curve is defined throughout the system $\hbar\frac{ \partial \Psi}{\partial \xi}(t,\xi)=\td{\mathcal{D}}(t,\xi)\Psi(t,\xi)$ with $\td{\mathcal{D}}(t,\xi)=-\frac{1}{\pi}\mathcal{D}(t,\xi)$):

\beq \text{Spectral Curve: } \lim_{\hbar\to 0}\,\, \det(Y I_2-\left(-\frac{1}{\pi}\right)\mathcal{D})=0\eeq

A straightforward computation gives the following characteristic polynomial (without taking the limit $\hbar \to 0$):
\beq \label{SpecP5}\pi^2Y^2=\alpha_0+\frac{\alpha_1}{\xi^2} +\frac{\alpha_2}{(\xi-1)^2} +\frac{\alpha_3}{\xi}+\frac{\alpha_4}{\xi-1}\eeq
with coefficients given by:
\bea \label{coeffs}
\alpha_0&=&-\frac{t^2}{4}\cr
\alpha_1&=&-\frac{1}{4}\theta_0^2\cr
\alpha_2&=&-\frac{1}{4}\theta_1^2\cr
\alpha_3&=& -\left(z+\frac{\theta_0}{2}\right)\left(t+2z+\theta_0+\theta_\infty\right)+ \frac{(z+\theta_0)}{y}\left(z+\frac{\theta_0+\theta_1+\theta_\infty}{2}\right)\cr
&&+zy\left(z+\frac{\theta_0-\theta_1+\theta_\infty}{2}\right)\cr
\alpha_4&=& (2z+\theta_0+\theta_\infty)(z+\frac{t}{2}+\frac{\theta_0}{2}) -yz\left(z+\frac{\theta_0-\theta_1+\theta_\infty}{2}\right)\cr
&&-\frac{(z+\theta_0)}{y}\left(z+\frac{\theta_0+\theta_1+\theta_\infty}{2}\right)\cr
\eea

It is worth mentioning that the function $\alpha_3+\alpha_4$ satisfies the following identity:
\beq \label{34} \alpha_3+\alpha_4=\frac{\theta_\infty}{2}t\eeq

When all monodromies are vanishing $\theta_0=\theta_1=\theta_\infty=0$ the spectral curve greatly simplifies into (using \eqref{reducedsigma}):
\beq \label{SpecCurvePainleve2}
\pi^2Y^2 = -\frac{t^2}{4} +\frac{\sigma(t)}{\xi}-\frac{\sigma(t)}{\xi-1}=-\frac{t^2\left(\xi^2-\xi+\frac{\sigma(t)}{t^2}\right)}{4\xi(\xi-1)}
\eeq

As presented earlier, the normalization used in the Lax pair is not directly suitable to get the universality results arising in matrix model and therefore we need to apply the change of variables \eqref{ChangeNormalization}. It leads to the spectral curve:
\beq  \label{SpecCurvePainleve}
\encadremath{Y^2 =  s^2 +\frac{\hat{\sigma}(s)}{\pi^2\xi}-\frac{\hat{\sigma}(s)}{\pi^2(\xi-1)}=\frac{s^2\xi^2-s^2\xi-\frac{1}{\pi^2}\hat{\sigma}(s)}{\xi(\xi-1)} }
\eeq 

In the next section we shall see that the spectral curve of a double-scaled hermitian matrix model always matches with this curve.

\subsubsection{Expansion at $\hbar= 0$}

In this paper we consider the asymptotic expansion in $\hbar$ of the $\sigma(t)$ function (solution of \eqref{Chazy}) that directly follows from the existence of an asymptotic expansion at large $t$ of the $\hbar=1$ Painlevé $5$ system proven in \cite{JMI} and the scaling argument presented later in section \ref{tdependance}. In a weaker sense, one could also simply consider a formal series expansion in $\hbar$. Anyway, a straightforward computation shows that the only non-trivial case (i.e. we want to avoid linear solution $\sigma(t)=\alpha t+\beta$ that always exists) corresponds to the following expansion:

\beq\label{conn} \sigma(t)\overset{\text{def}}{=}\sum_{k=0}^\infty \pi^{2k}\sigma_{k}\hbar^{2k}t^{2-2k}\eeq

We stress here that since the differential equation \eqref{Chazy} only involves $\hbar^2$ but not directly $\hbar$ we automatically get the existence of an asymptotic series in $\hbar^2$ rather than $\hbar$. Inserting the series expansion into the differential equation gives the following recursion:
\bea \label{recursion} \sigma_0&=&\frac{1}{16}\,\,,\,\, \sigma_1=\frac{1}{4}\,\,,\,\, \sigma_2=1\,\,,\,\, \sigma_3=40\,\,,\,\, \sigma_4=4192\cr
\sigma_{k}&=&16\sum_{j=0}^{k-1}(2-2j)(1-2j)(4+2j-2k)(3+2j-2k)\sigma_{j}\sigma_{k-j-1}\cr
&&-16\sum_{i=1}^{k-1}(2i-1)(1-2k+2i)\sigma_{i}\sigma_{k-i}\cr
&&-64\sum_{i=1}^{k-1}\sum_{j=0}^i(2-2j)(2-2i+2j)(1-2k+2i)\sigma_{j}\sigma_{i-j}\sigma_{k-i}\cr
\eea
In particular, we observe that this recursive form implies that \textbf{$\forall k\geq2$ : $\sigma_{k}$ is an integer} which is not trivial from the differential equation. This surprising fact may have some interests in combinatorics as it is often the case with matrix models but so far we do not have any real application in relation with this observation. The proof of \eqref{recursion} is presented in appendix \ref{AppendixA}.
From these results and the correspondence \eqref{ChangeNormalization} it is easy to deduce the asymptotic expansion of $\hat{\sigma}(s)$:
\beq \hat{\sigma}(s)=-\pi^2\sum_{k=0}^\infty (-1)^k\sigma_{k}(2s)^{2-2k}\hbar^{2k}=-\pi^2\left(\frac{s^2}{4}-\frac{1}{4}\hbar^2+\frac{\hbar^4}{4s^2}\right)+O(\hbar^6) \eeq

\subsubsection{\label{tdependance} Time rescaling and alternative interpretation of $\hbar$ }

Another interpretation or natural way to introduce the $\hbar$ factor in the Lax system is to observe the following: if we define:
\bea (\td{\theta}_0,\td{\theta}_1,\td{\theta}_\infty)&=&\left(\frac{\theta_0}{\hbar},\frac{\theta_1}{\hbar},\frac{\theta_\infty}{\hbar}\right)\cr
\left(\td{\sigma}(t),\td{z}(t),\td{y}(t),\td{u}(t)\right)&=&\left(\frac{1}{\hbar^2}\sigma\left(\frac{t}{\hbar}\right),\frac{1}{\hbar}z\left(\frac{t}{\hbar}\right),y\left(\frac{t}{\hbar}\right),u\left(\frac{t}{\hbar} \right)\right)\cr
\td{\Psi}(t,\xi)&=&\Psi(\frac{t}{\hbar},\xi)\cr
\eea
then it is straightforward to observe that the Lax equations satisfied by $\td{\Psi}(t,\xi)$ are $\hbar$-independent:

\bea -\pi\frac{ \partial \td{\Psi}}{\partial \xi}(t,\xi) &=&\left( \frac{t}{2}\sigma_3 +\frac{\td{A}_0}{\xi}+\frac{\td{A}_1}{\xi-1}\right) \cr
-\pi\frac{ \partial \td{\Psi}}{\partial t}(t,\xi) &=&\left( \frac{\xi}{2}\sigma_3 +\frac{1}{t}\left(\td{A}_0+\td{A}_1+\frac{\td{\theta}_\infty}{2}\sigma_3\right)\right) \td{\Psi}(t,\xi)\cr
 \eea
with the matrices $\td{A}_0$ and $\td{A}_1$ identical to $A_0$ and $A_1$ except that all function $z,y,u$ have to be changed for their tilde counterpart. Consequently, the Painlevé $5$ equation satisfied by $\td{\sigma}$ also becomes $\hbar$-invariant (and recover the more standard one):

\bea
\label{P55}
&&-\left( \pi t\ddot{\td{\sigma}}\right)^2+\left(\td{\sigma}+2(\dot{\td{\sigma}})^2-(2\td{\theta}_0+\td{\theta}_\infty+ t)\dot{\td{\sigma}}\right)^2=\cr
&&4\dot{\td{\sigma}}(\dot{\td{\sigma}}-\td{\theta}_0)\left(\dot{\td{\sigma}}- \frac{\td{\theta}_0-\td{\theta}_1+\td{\theta}_\infty}{2}\right)\left(\dot{\td{\sigma}}- \frac{\td{\theta}_0+\td{\theta}_1+\td{\theta}_\infty}{2}\right) 
\eea

thus ensuring also that $\td{z}$, $\td{y}$ and $\td{u}$ are $\hbar$-invariant too.

\medskip

In the case of vanishing monodromies, the previous observation has important consequences regarding the asymptotic expansions of both $\sigma(t)$ and $\Psi(t,\xi)$ (and consequently $y(t),z(t)$ and $u(t)$). Indeed, from the previous remark, we observe that the $\hbar$ asymptotic expansion of these functions can be recast in a large $t$ expansion for the tilde functions. Since both times are connected through $\td{t}=\frac{t}{\hbar}$, then we immediately get the following theorem:

\medskip

\textbf{The existence of a $\hbar$ asymptotic series for quantities arising in the $\hbar$ Lax pair formalism is equivalent to the existence of a $\frac{1}{t}$ asymptotic series for the corresponding quantities arising in the $\hbar=1$ Lax pair formalism}.

\medskip

For example, the existence of an asymptotic series $\td{\sigma}(t)=\underset{k=0}{\overset{\infty}{\sum}} \frac{c_k}{t^{2k-2}}$ is equivalent to $\sigma(t)=\underset{k=0}{\overset{\infty}{\sum}} c_{k}\hbar^{2k}t^{2-2k}$ with the same numbers $c_k$. As mentioned earlier since the existence of a series expansion at large $t$ has been proved in the $\hbar=1$ Lax pair \cite{JMI}, it automatically justifies the existence of our series expansion \eqref{conn}.

\subsection{The $\hbar$ expansion of the $\tau$-function}

From the existence of the $\hbar$ asymptotic expansion of the $\hat{\sigma}(s)$ function, we can deduce the asymptotic expansion of the corresponding $\tau$-function. With the introduction of the $\hbar$ parameter (and the corresponding $\pi$ factor attached to it), the definition of the $\tau$-function is modified into:
\beq \pi^2\hbar^2\log \hat{\tau}(s)=\int^s\frac{\hat{\sigma}(u)}{u}du \,\Leftrightarrow \hat{\sigma}(s)=\hbar^2\pi^2 s\frac{d}{ds}\log \hat{\tau}(s)\eeq
Note that the $\tau$ function is only defined up to a multiplicative constant (as usual in integrable system) which corresponds to choosing the lower bound of the integral in the previous equation. We therefore take term by term integrals discarding the undefined coefficients in accordance to the expected singular behavior at $s=0$ corresponding to the merging of the two edges of the gap interval. Therefore $\log \hat{\tau}(s)$ has an asymptotic expansion of the form:
\beq \label{taudef}\log \hat{\tau}(s)\overset{\text{def}}{=}\sum_{k=0}^\infty \hat{\tau}_k(s) \hbar^{2k-2}
\eeq
with \beq \forall \, k\neq 1\,\,:\,\,\hat{\tau}_k(s)=(-1)^{k-1}\frac{(2s)^{2-2k}}{2-2k}\sigma_k \,\,\text{ and }\,\, \hat{\tau}_1(s)=\sigma_1 \ln s\eeq
The first orders are:
\beq \label{TauCoeff}\encadremath{\log \hat{\tau}(s)=-\frac{s^2}{8\hbar^2}+\frac{1}{4}\ln s+\frac{\hbar^2}{8s^2}-\frac{5\hbar^4}{8s^4}+\frac{131\hbar^6}{12s^6}-\frac{6375\hbar^8}{16s^8}+O(\hbar^{10}) }\eeq

The purpose of the article is to prove using the topological recursion that this $\tau$-function has the same asymptotic expansion (to all orders) as the one naturally occurring in matrix models.

\section{Matrix models side}

\subsection{Gap probability}

Consider a $N\times N$ hermitian random matrix $M$ with a probability law given by a Boltzmann weight with a potential $V(x)$, which we assume polynomial and bounded from below, i.e.:
\beq
d\mu(M)=\frac{1}{\tilde Z}\,\,dM \,\,\e^{-\frac{N}{T}\,\Tr V(M)}
\eeq
where
\beq
\tilde Z = \int_{H_N} dM  \,\,\e^{-\frac{N}{T}\,\Tr V(M)}
\eeq
The induced law for the eigenvalues of $M$ is obtained by diagonalizing
$$
M = U\,\Lambda \, U^{\dagger}
$$
where $U\in U(N)$ and $\Lambda={\rm diag}(\lambda_1,\lambda_2,\dots,\lambda_N)$.
One gets (see \cite{MehtaBook}):
\beq
d\mu(\Lambda) = \frac{1}{Z}\,\Delta(\lambda)^2\,\,\prod_{i=1}^N \ee{-\frac{N}{T}\,V(\lambda_i)}\,d\lambda_i
\eeq
with the Vandermonde determinant:
$$
\Delta(\lambda) = \prod_{1\leq i<j\leq N}\,(\lambda_j-\lambda_i)
$$
and
$$
Z = \int_{\mathbb R^N} \Delta(\lambda)^2\,\,\underset{i=1}{\overset{N}{\prod}} \ee{-\frac{N}{T}\,V(\lambda_i)}\,d\lambda_i
$$

The {\bf gap probability} is the probability that there are no eigenvalues inside an interval $[x_0,x_1]$ within the bulk of the limiting distribution. In the matrix models terminology, this is called having ``hard edges" at $x_0$ and $x_1$, i.e. an integration contour for eigenvalues which ends at a point where the measure is normally not vanishing. In other words, it is given by:
\beq p_2(x_0,x_1)=\frac{ \int_{\left(\mathbb{R}\setminus\left[x_0,x_1\right]\right)^N} \Delta(\lambda)^2\,\,\underset{i=1}{\overset{N}{\prod}} \ee{-\frac{N}{T}\,V(\lambda_i)}\,d\lambda_i} {\int_{\mathbb R^N} \Delta(\lambda)^2\,\,\underset{i=1}{\overset{N}{\prod}} \ee{-\frac{N}{T}\,V(\lambda_i)}\,d\lambda_i}\eeq
 
\subsection{Notations and loop equations}

Loop equations for matrix models with hard edges have been developed in \cite{HardEdgess}.They consist in finding algebraic relations between the following expectation values:
\bea W_1(x)&=&\left<\sum_{i=1}^{N} \frac{1}{x-\lambda_i}\right>\cr
W_p(x_1,\dots,x_p)&=&\left< \sum_{i_1,\dots i_p} \frac{1}{x_1-\lambda_{i_1}}\dots\frac{1}{x_p-\lambda_{i_p}}\right>_c \cr
P_p(x_1,x_2,\dots,x_p)&=& \left< \sum_{i_1,\dots i_p}\frac{V'(x_1)-V'(\lambda_{i_1})}{x_1-\lambda_{i_1}}\frac{1}{x_2-\lambda_{i_2}}\dots\frac{1}{x_p-\lambda_{i_p}}\right>_c
\cr
\eea
$W_1(x)$ is the expectation value of the resolvent, also called one-point function, it is the Stieljes transform of the density of eigenvalues:
\beq
W_1(x)=\int_{{\rm supp}.\,d\rho} \frac{d\rho(z) }{x-z}
\eeq
where
\beq
d\rho(z) = \left<\sum_{i=1}^{N} \delta(z-\lambda_i)\right>\,\,dz.
\eeq

The functions $W_p(x_1,\dots,x_p)$ are called $p$-point connected correlation functions (the subscript $<.>_c$ means cumulant).

The function $P_p$ is defined only for technical intermediate purposes, it has the property to be a polynomial function of its first argument.
Loop equations are traditionally obtained from integration by parts but in the case of hard edges this method is tricky since we have to get into account the boundary contributions if we compute $\left<\underset{i=1}{\overset{N}{\sum}}\frac{\partial}{\partial \lambda_i}\left(\frac{1}{x-\lambda_i}\right)\right>$. A possible way to avoid those terms which we develop here is to properly adapt the numerator of the previous quantity so that the boundary terms vanish. We get:
\bea
0&=& \sum_{i=1}^N \int_{\mathbb R\setminus[x_0,x_1]} d\lambda_1\dots d\lambda_N \frac{\partial}{\partial \lambda_i} \left( \frac{(x_0-\lambda_i)(x_1-\lambda_i)}{x-\lambda_i}\,\,\prod_{k<j} (\lambda_k-\lambda_j)^2\,\,\prod_k \ee{-\frac{N}{T} V(\lambda_k)}\right)  \cr
&=& \sum_{i=1}^N \int_{\mathbb R\setminus[x_0,x_1]} d\lambda_1\dots d\lambda_N  \prod_{k<j} (\lambda_k-\lambda_j)^2\,\,\prod_k \ee{-\frac{N}{T} V(\lambda_k)}\,\,\Big( \frac{(x_0-\lambda_i)(x_1-\lambda_i)}{(x-\lambda_i)^2} \cr
&& + \sum_{j\neq i}\frac{2(x_0-\lambda_i)(x_1-\lambda_i)}{(x-\lambda_i)(\lambda_i-\lambda_j)} - \frac{N}{T}\,\frac{V'(\lambda_i)(x_0-\lambda_i)(x_1-\lambda_i)}{x-\lambda_i}+\frac{2\lambda_i-x_0-x_1}{x-\lambda_i}\,\,\Big)  \cr
\eea

Using the decomposition $(x_0-\lambda_i)(x_1-\lambda_i)=(x-x_0)(x-x_1)+(x_0+x_1-2x)(x-\lambda_i)+(x-\lambda_i)^2$,we get (see \cite{HardEdgess}):

\bea \label{spec} &&W_1(x)^2+W_2(x,x)=\frac{N}{T}V'(x)W_1(x)-\frac{N}{T}P_1(x) \cr
&& +\frac{1}{(x-x_0)(x-x_1)}\,\left(N^2-\frac{N}{T}<\Tr (M+x-x_0-x_1)V'(M)>\right) 
\eea
Notice that without hard edges we would have $<\Tr V'(M)>=0$ by translation invariance of the eigenvalues, and $<\Tr M V'(M)>=N T$ by dilatation invariance, which are both broken by the presence of two hard edges.

\subsection{Large $N$ topological expansion}

Let us assume that our matrix model has a large $N$ expansion, i.e. all correlation functions have a $\frac{1}{N^2}$ expansion of the form:
\beq
W_1(x) = \sum_{g=0}^\infty \left(\frac{N}{T}\right)^{1-2g} W_1^{(g)}(x)
\eeq
\beq
W_p(x_1,\dots,x_p) = \sum_{g=0}^\infty \left(\frac{N}{T}\right)^{2-2g-p} W_p^{(g)}(x_1,\dots,x_p).
\eeq
\beq
P_p(x_1,\dots,x_p) = \sum_{g=0}^\infty \left(\frac{N}{T}\right)^{2-2g-p} P_p^{(g)}(x_1,\dots,x_p).
\eeq
\beq \ln Z= \sum_{g=0}^\infty \left(\frac{N}{T}\right)^{2-2g} F^{(g)}
\eeq
where the coefficients $W_p^{(g)}$ and $F^{(g)}$ are functionals of the potential $V(x)$ and functions of $x_0$ and $x_1$.

\smallskip

This assumption has been proved to hold for many potentials (typically convex potentials $V(x)$, but also more general cases) \cite{Guionnet,Pastur,Ambjorn, BG1}, and is also known to be wrong for some potentials (typically potentials with several wells), where the correlation functions have oscillatory behaviors at large $N$ \cite{Anomaly,Anomaly2, BG2}.

Here, we shall not debate on the specific conditions required to prove the existence of such an expansion, we shall assume that our matrix model potential is such that such an expansion exists.

\subsection{Getting the spectral curve and the double-scaling limit}

The spectral curve is defined as the leading order of the first loop equation and is classically presented with the function, $Y(x)=W_1^{(0)}(x)-\frac{V'(x)}{2}$. The spectral curve is thus:
\beq \label{rrr}
Y^2(x) =\frac{V'^2(x)}{4} -P_1^{(0)}(x) + \frac{a_0}{x-x_0}+\frac{a_1}{x-x_1}
\eeq
where $a_0$ and $a_1$ are coefficients to be determined with additional considerations. In fact a close analysis shows that they are formally expressed as:
\bea a_0&=& \frac{-1+\underset{N\to\infty}{\lim}<\frac{1}{NT}<\Tr(MV'(M))>-x_1\underset{N\to\infty}{\lim} \frac{1}{NT}<\Tr(V'(M))>}{x_1-x_0}\cr
a_1&=&\frac{1-\underset{N\to\infty}{\lim}<\frac{1}{NT}<\Tr(MV'(M))>+x_0\underset{N\to\infty}{\lim} \frac{1}{NT}<\Tr(V'(M))>}{x_1-x_0}\cr 
\eea
leading in particular to:
\beq a_0+a_1=-\frac{1}{T}\underset{N\to\infty}{\lim} \frac{1}{N}<\Tr(V'(M))>\eeq

We observe that when $x_0\to x_1$ the hard edges disappear, and the spectral curve should be $Y^2(x)= \frac{V'(x)^2}{4}-P(x)$, i.e. we expect the coefficients $a_0$ and $a_1$ to be vanishing in the limit $x_0\to x_1$.We shall make the assumption that they vanish linearly in $O(x_0-x_1)$ which is coherent with the fact that $\frac{1}{N}<\Tr(V'(M))>$ is expected to vanish linearly when $x_0\to x_1$. It also corresponds to the standard double scaling limit in the bulk for which we expect universality.

We want to study a \textbf{double scaling limit around $x_0$}. It means that we should shrink the gap interval $[x_0,x_1]$ simultaneously with the increase of $N$. Therefore we need to introduce the following scaling:
\bea \label{scaling limit} x(\xi)&=&x_0+\xi (x_1-x_0)\cr
x_1-x_0&=&\frac{s}{N}\cr
&N\to \infty \, ,\, x_0 \text{ fixed}&
\eea
This scaling limit corresponds to focusing around the point $x_0$ when $N$ goes to infinity and thus to recover universality which is known only to apply for local statistics around a fixed point (here a regular bulk point) of the limiting eigenvalues density. Concretely, it transforms the global variable $x$ to the local variable $\xi$. The parameter $s$ introduced in the scaling limit \eqref{scaling limit} controls the renormalized size of the gap of eigenvalues. This interpretation makes it obvious that it should be trivially related to the time parameter $t$ arising in the Painlevé $5$ formalism.
Eventually, we remind the reader that standard results regarding universality in the bulk (as the gap probability presented in section \ref{section1}) are expressed with the requirement that the density of eigenvalues is taken at a normalized point (density set to $1$). Here we deal with the spectral curve that also has to be normalized properly. In our case, it corresponds to defining $y(\xi)$ as the following rescaling:
\beq \frac{Y(x)}{\td{Y}(x_0)}dx=\frac{1}{N}y(\xi)d\xi \,\, \text{ with } \td{Y}(x_0)=\frac{V'(x_0)^2}{4}-P(x_0)\eeq
$\td{Y}(x_0)$ is required to normalize properly the spectral curve and thus set the density at $x_0$ to $1$. The function $P(x_0)$ is the corresponding value of $P_1^{(0)}(x)$ when the hard edges are removed (i.e. $\td{Y}(x_0)$ defines the spectral curve for the limiting curve without hard edges). Since we have assumed that $x_0$ is a regular bulk point, then $\rho(x_0)>0$, i.e. $\td{Y}(x_0)\neq 0$ ensuring that the former definition makes sense. To sum up, we have detailed here how to pass from global variables $(x,Y)$ (for which the relation $Y(x)=0$ describe the global spectral curve, hence the global eigenvalues distribution) to local (focused around a regular bulk point $x_0$) variables $(\xi,y)$ corresponding to the double scaling limit around $x_0$. The local (double-scaled) spectral curve is then obtained by the $N\to \infty$ limit of \eqref{rrr} in the local variables. We find: 
\beq \label{SpectralCurve Simplified}\encadremath{y^2(\xi)=s^2+\frac{b_0(s)}{\xi}+\frac{b_1(s)}{\xi-1} } \eeq
which is exactly of the same form as the spectral curve arising on the Painlevé $5$ side \eqref{SpecCurvePainleve} with the identification $t=s$. From \cite{EO}, we know that the topological recursion commutes with such limit and therefore we expect the asymptotic expansion of the limiting partition function (which is known to be the $\tau$-function) to be given by the free energies (also called symplectic invariants) $F^{(g)}$ of the curve \eqref{SpectralCurve Simplified}. In other words, we expect:
\beq F^{(g)}(s)=\hat{\tau}_k(s)\,\,\, \forall \, k\geq\,0\eeq
where $\hat{\tau}_k(s)$ are given in \eqref{taudef}. Moreover, the correlation functions $W_n^{(g)}(x_1,\dots,x_n)$ are also known to have a scaling limit and therefore will define double-scaled functions $W_n^{(g)}(\xi_1,\dots,\xi_n)$ that can be computed with the topological recursion applied to the new spectral curve. These functions will also be proved to match the ones built from the Lax pair presented earlier.

Note that so far, the matrix model analysis does not fully determine the spectral curve since the constant $b_0(s)$ and $b_1(s)$ (coming from the scaling of $a_0$ and $a_1$ that were unknown too) are not determined with the loop equations method. In order to do so, we need two other conditions, naturally arising in the matrix models perspective that we discuss in the next section.

\subsection{The genus zero assumption}

We shall now explain how to determine the parameters $b_0(s)$ and $b_1(s)$.

Classically in matrix models, the loop equations do not completely determine the spectral curve and additional information must be inserted in order to do so. For convergent matrix models, the use of potential theory gives a complete way to determine the whole spectral curve as explained for example in \cite{BleherNotes}. However the method used there is tedious and even if the limiting eigenvalue distribution is proved to exist and satisfy a variational problem, finding explicitly the solution to this problem is in general quite involved. Fortunately, in many cases some heuristic arguments give us some clear insights about what to be done and in our present situation this can be carried on in full details. Indeed, in our case, we have forced the absence of eigenvalues inside $[x_0,x_1]$ by inserting two hard edges at $x=x_0$ and $x=x_1$. Classically, inserting hard edges in an hermitian matrix model forces the support of the eigenvalues distribution to end at those hard edges. Since we have assumed that $x_0$ is a regular point in the bulk, and because the rescaled $\xi$ variable zooms around $x_0$, we expect the cuts in the $\xi$ variable to extend from $-\infty$ to $0$ and then from $1$ to $\infty$. In general inserting hard edges in $x=x_0$ and $x=x_1$ does not necessarily mean that there will be no eigenvalues inside $[x_0,x_1]$. Therefore if we did not look at the gap probability, we would expect another cut inside $[0,1]$ in the $\xi$ variable. This would mean that the function $y^2(\xi)$ described in \eqref{SpectralCurve Simplified} should in general have two distinct zeros inside $[0,1]$. But here, since we are dealing with the gap probability, we have imposed the absence of eigenvalues inside $[0,1]$ by construction and thus the only possible way to achieve this situation is the degenerate case happening when the two inner roots collapse to a double zero. We illustrate the discussion with the following picture:

\begin{center}
\includegraphics[width=8cm]{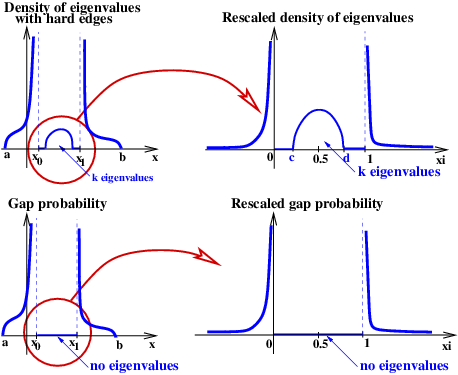}
\end{center}
\textit{\underline{Fig.1}: Rescaling around a regular point of the bulk in the case of hard edges with or without imposing the gap condition}

\medskip

Regarding the previous argument, we expect the matrix model rescaled spectral curve to be of the form:
\beq \label{specmod} y^2(\xi)=\frac{s^2(\xi-a)^2}{\xi(\xi-1)}\,\,,\,\, a\in [0,1]\eeq

Moreover, if we want to avoid the tunnel effect between the two sides we need to impose equilibrium of the chemical potentials on both sides. In matrix models, this leads to the condition:
\beq \int_0^1 y(\xi)d\xi=0\eeq
Since $y(\xi)$ is given by \eqref{specmod}, the symmetry of the integral only leads to one solution: $a=\frac{1}{2}$ and thus we get on the matrix model side the spectral curve:
\beq \label{SpecCurveMM} \encadremath{y^2(\xi)=\frac{s^2\left(\xi-\frac{1}{2}\right)^2}{\xi(\xi-1)} }\eeq 
In particular, matching \eqref{SpecCurveMM} with \eqref{SpecCurvePainleve} leads to the fact that the leading order of the $\hat{\sigma}(s)$ function must be given by $\hat{\sigma}_0(s)=-\frac{\pi^2s^2}{4}$ which is indeed the case.

Hence, we conclude that \textbf{the two spectral curves arising from the matrix model and the integrable system side match.}

\medskip

\underline{Note}: In the end we observe that the spectral curve arising on the matrix model side is symmetric about $\xi=\frac{1}{2}$. Although we started from a possibly non-symmetric eigenvalues distribution (there is a priori no reason that $x_0$ is a center of symmetry of the eigenvalues distribution without hard edges), the double scaling limit focuses around $x_0$ so that in the leading order, the general shape of the eigenvalue distribution has no influence on the local spectral curve. This is not surprising since local statistics are expected to be universal and using directly the knowledge of universality, we could have directly guessed that the spectral curve had to be symmetric about $\xi=\frac{1}{2}$.

\subsection{Parametrization in the Zhukovsky variable}

In order to apply the topological recursion, we need to choose a proper parametrization of the spectral curve \eqref{SpecCurveMM} around the branchpoints. Here the hyperelliptic equation defines a genus zero Riemann surface with two sheets and therefore a global parametrization with a variable living on $\bar{\mathbb{C}}$ exists. The branchpoints are located at $\xi=0$ and $\xi=1$ giving a standard Zhukovsky parametrization (Cf. \cite{OM}) given by:
\bea \label{zhu}\xi(z)&=&\frac{1}{2}+\frac{1}{4}\left(z+\frac{1}{z}\right)\cr
 y(z)&=&\frac{ s(z^2+1)}{(z^2-1)}\cr
\eea
We will apply the topological recursion developed by in \cite{EO} to this curve in a following section and we shall see that the symplectic invariants $F^{(g)}$ match the series expansion of the $\tau$-function arising from Painlevé $5$.

\section{Topological recursion and invariants}

In the previous section, we have been able to identify the spectral curves coming from the matrix model problem \eqref{SpecCurveMM} and the spectral curve (as defined by \cite{BE}) arising in the Painlevé $5$ Lax pair \eqref{SpecP5}. This tells us that the first correlation function $W_1^{(0)}(\xi)$ (from which $y(\xi)$ is directly defined) are the same on both sides. In this section, we want to prove that all correlation functions are the same on both sides at any order (formally in powers of $\hbar$ or $\frac{1}{N}$). Unfortunately, this result is not automatic even if the initial spectral curves are the same. Indeed, the formalism developed on both sides to define correlation functions is quite different: on the matrix model side, correlations functions $W_n^{(g)}(\xi_1,\dots,\xi_n)$ can be obtained by the topological recursion as explained in \cite{EO}. On the integrable system side, some correlation functions can be defined with determinantal formulas as presented in \cite{BE}. In order to be sure that both quantities match, we need to prove that a few technical conditions are realized. In general, this is achieved by studying in details the pole structure on the integrable system side in order to show that some possible unwanted poles do not arise in the formulas. The key ingredient to get such results is to use the second equation of the Lax pair that so far has not been fully used. This is what we will do here after reviewing rapidly the formalisms on both sides.  

\subsection{The topological recursion of \cite{EO}}

In \cite{EO}, was introduced a recursive way to construct the corresponding correlation functions $W_n^{(g)}$ and $F^{(g)}$ from any spectral curve. In our case, the curve is of genus zero so that:
\beq W_2^{(0)}(\xi_1,\xi_2)=\frac{d \xi_1 \otimes d\xi_2}{(\xi_1-\xi_2)^2}\eeq

Since the curve is of genus $0$, there exists a global parametrization of the curve with the introduction of the Zhukovsky variable given in \eqref{zhu}. With this parametrization, the one-form $yd\xi$ is a meromorphic form in $z$ and the branchpoints (zeros of $d\xi$) are $z=\pm 1$. Since the curve is hyperelliptic, the conjugate point around the branchpoints are defined globally and are obtained by $\bar{z}=\frac{1}{z}$. We define the following forms:
\bea \omega(z)&=&(y(z)-y(\bar{z}))d\xi(z)\cr
 B(z,z')&=& \frac{dz\otimes dz'}{(z-z')^2}\cr
 \mathcal{K}(z_0,z)&=&\frac{1}{2}\int_z^{\bar{z}}B(s,z)ds=\frac{zdz}{z^2-z_0^2}\cr
 \Phi(z)&=& \int^z yd\xi\cr
 \eea
Then, the correlation functions are defined by recursion $I=\{z_1,\dots,z_n\}$ and $A=\{a,b\}$:
\bea \label{Toprec} W_{n+1}^{(g)}(z_1,\dots,z_n)&=& \sum_{a_i\in A } \Res_{z\to a_i} \frac{\mathcal{K}(z_1,z)}{\omega(q)}\Big( W_{n+2}^{(g-1)}(z,\bar{z},p_{I})\cr
&&+\sum_{m=0}^g \sum'_{I_1\sqcup I_2=I} W_{|I_1|+1}^{(m)}(z,z_{I_1}) W_{|I_2|+1}^{(g-m)}(\bar{z},z_{I_2}) \Big)
\eea 

In particular, from the topological recursion, it is well known that the correlation functions satisfy the loop equations and may only have poles at the branchpoints of the curve, i.e. $z=\pm1$. In fact, the structure of poles at the branchpoints is characteristic of the solution of the loop equations. Indeed, it is known that although there are many possible solutions of the loop equations, there is only a unique one with poles only at the branchpoints of the spectral curve. For reasons that we will not develop here, these particular solutions are the ones of interest in matrix models and thus appear naturally as the solution in the topological recursion. In the same spirit we can also define the symplectic invariants as:
\beq \forall\,g\geq 2\, , \qquad F^{(g)}=W_0^{(g)}= \frac{1}{2-2g}\sum_{a_i\in A } \Res_{z\to a_i} \Phi(z)W_1^{(g)}(z)\eeq
They are invariant under any symplectic change of the spectral curve ($(\xi,y)\to (\td{\xi},\td{y})$ such that $d\xi\wedge dy=d\td{\xi} \wedge d\td{y}$). In the next sections we will prove the following:
\begin{theorem} \label{MainTheorem}
The symplectic invariants $F^{(g)}$ computed from the spectral curve obtained in the double-scaling limit of an hermitian matrix model around a regular point in the bulk are identical to the series expansion of $\ln \tau$ arising from the Painlevé $5$ Lax pair (defined in \eqref{TauCoeff}).
\end{theorem}  

\subsection{Computation of the first orders of the topological recursion}

In this section, we implement the topological recursion in order to get the first correlation functions and free energies of the curve \eqref{SpecCurveMM}. In particular, we will find agreement with \ref{MainTheorem}. The computation from \eqref{Toprec} can be carried out with a symbolic computation software and thus we will only give our results there. We warn the reader that even with a symbolic computation software, the computations rapidly become difficult and time consuming. In order to have rational functions, we use the Zhukovsky parametrization:
\bea
\xi(z)&=& \frac{1}{2}+\frac{1}{4}\left(z+\frac{1}{z}\right)\cr
y(z)&=&\frac{s(z^2+1)}{(z^2-1)}\cr
\eea
Therefore, the fundamental differential is given by:
\beq \omega(z)=y(z)dx(z)=\frac{s(z^2+1)}{4 z^2}dz\eeq
and the fundamental bi-differential that starts the recursion is (it is the same for every genus zero curve):
\beq B(z_1,z_2)= W_2^{(0)}(z_1,z_2)=\frac{dz_1 \, dz_2}{(z_1-z_2)^2}\eeq 
Note that the spectral curve is a little singular from traditional genus $0$ curves. Indeed, if the $x$ function is standard with two branchpoints at $z=\pm 1$ and a conjugate local coordinate given by $\bar{z}=\frac{1}{z}$, the $y$ function does not have simple zeros at those branchpoints but rather some simple poles there. Fortunately, this does not affect the standard topological recursion which can be carried out by the standard formulas developed in \cite{EO}. The next step is to compute the recursion kernel function $\mathcal{K}(z_0,z)$. We find: 
\beq \frac{\mathcal{K}(z_0,z)}{\omega(z)}=\frac{(z^2-1)z^2}{(z_0z-1)(z-z_0)(z^2+1) s}\eeq
From there, the computation of the topological recursion can be performed by taking residues at $z=\pm 1$. We only mention here the results of the computation:

\bea \label{Computations}W_n^{(0)}(z_1,\dots,z_n)&=&0 \,\, \forall n\geq 2\cr
W_1^{(1)}(z)&=&-\frac{(z^2+1)}{2s(z^2-1)^2}\cr
W_2^{(1)}(z_1,z_2)&=&\frac{}{2s^2(z_1-1)^2(z_2-1)^2}+\frac{1}{2s^2(z_1+1)^2(z_2+1)^2}\cr
W_1^{(2)}(z)&=&-\frac{(z^2+1)(z^4-11z^2+1)}{2s^3(z^2-1)^4}\cr
W_3^{(1)}(z_1,z_2,z_3)&=&\frac{2}{s^3}\left(\frac{1}{(z_1-1)^2(z_2-1)^2(z_3-1)^2}+ \frac{1}{(z_1+1)^2(z_2+1)^2(z_3+1)^2}\right)\cr
W_1^{(3)}(z)&=&\frac{(z^2+1)(5z^8-47z^6+309z^4-47z^2+5)}{s^5(z^2-1)^6}\cr
W_4^{(1)}(z_1,\dots,z_4)&=&\frac{12}{s^4}\left(\frac{1}{(z_1-1)^2\dots(z_4-1)^2}+\frac{1}{(z_1+1)^2\dots(z_4+1)^2}\right)\cr
W_1^{(4)}(z)&=&-\frac{(262z^{12}-2841z^{10}+15756z^8-81479z^6+15756z^4-2841z^2+262)}{2s^7(z^2-1)^8}\cr
W_5^{(1)}(z_1,\dots,z_5)&=&\frac{96}{s^5}\left(\frac{1}{(z_1-1)^2\dots(z_5-1)^2}+\frac{1}{(z_1+1)^2\dots(z_5+1)^2}\right)\cr
W_1^{(5)}(z)&=&\frac{5(z^2+1)z^8}{s^9(z^2-1)^{10}}\Big(1315(z^8+z^{-8})-16588(z^6+z^{-6})+102133(z^4+z^{-4})\cr
&&-434810(z^2+z^{-2})+1646095\Big)\cr
\eea

We computed all correlation functions necessary to get $F^{(5)}$ we find:

\bea 
F^{(2)}&=&\frac{1}{8s^2}&  \cr
F^{(3)}&=&-\frac{5}{8s^4}  \cr
F^{(4)}&=&\frac{131}{12s^6}  \cr
F^{(5)}&=&-\frac{6575}{16 s^8}\cr
\eea
Specific computations (that can be adapted directly from \cite{EO} in our case) also give the two leading free energies:
\bea F^{(0)}&=&-\frac{s^2}{8}\cr
F^{(1)}&=&\frac{1}{4}\ln s\cr
\eea
that is to say that the $\tau$-function associated to the matrix model has a formal series expansion of the form:
\beq \ln \tau=\sum_{g=0}^\infty F^{(g)}\hbar^{2g-2}=-\frac{s^2}{8\hbar^2}+\frac{1}{4}\ln s+\frac{\hbar^2}{8s^2}-\frac{5\hbar^4}{8s^4}+\frac{131\hbar^6}{12s^6}-\frac{6575\hbar^8}{16 s^8}+O(\hbar^{10})\eeq

In other words, \textbf{up to the fifth order we recover exactly the asymptotic series of the $\tau$-function arising from the Painlevé $5$ side \eqref{taudef}}

We shall see in the next section how to prove the identity to all orders using uniqueness of the solution of the loop equations.

\section{Determinantal formulas on the integrable system side}

In \cite{BE} is presented a natural way to build ``correlation functions" from any $2\times 2$ Lax pair. Indeed, from any Lax pair of the form \eqref{Lax}, they introduce:
\beq \hbar\frac{\partial}{\partial t} \Psi(t,\xi)= \mathcal{D}(t,\xi) \Psi(t,\xi)=\begin{pmatrix}
a&b \\
c&d
\end{pmatrix}\Psi(t,\xi)\,\,,\,\,
\Psi(t,\xi)=\begin{pmatrix}
\psi(t,\xi)&\phi(t,\xi) \\
\td{\psi}(t,\xi)&\td{\phi}(t,\xi)
\end{pmatrix}
\,\,,\det(\Psi)=1\eeq
with $a,b,c,d$ rational functions of $\xi$ and $a+d=0$ (Traceless form of the Lax pair). Then one can define the Christoffel-Darboux kernel by:
\beq \label{KK} K(\xi_1,\xi_2)=\frac{\psi(t,\xi_1)\td{\phi}(t,\xi_2)- \td{\psi}(t,\xi_1)\phi(t,\xi_2)}{\xi_1-\xi_2}\eeq
Then the correlations functions are defined by:
\beq W_1(\xi)=\lim_{\xi_2\to \xi}K(\xi,\xi_2)=\psi'(t,\xi)\td{\phi}(t,\xi)-\td{\psi}(t,\xi)\phi'(t,\xi)\eeq
and recursively:
\beq \label{deter} W_n(\xi_1,\dots,\xi_n)=-\frac{\delta_{n,2}}{(\xi_1-\xi_2)^2}+(-1)^{n+1}\sum_{\tau \text{ cycles}}\prod_{i=1}^nK(\xi_i,\xi_{\tau(i)})\eeq
The authors are able to show that these functions satisfy some loop equations similar to the ones found in matrix models $(L=\xi_1,\dots,\xi_n)$:
\bea \label{tyuu}-P_{n+1}(\xi;L)&=&W_{n+2}(\xi,\xi,L)+\sum_{J \subset L}W_{1+|J|}(\xi;J)W_{1+n-|J|}(\xi;L\setminus J)\cr
&&+ \sum_{j=1}^n\frac{d}{d \xi_j} \frac{W_n(\xi,L\setminus \{\xi_j\})-W(\xi_j,L\setminus \{\xi_j\})}{\xi-\xi_j}\cr
\eea
with $P_{n+1}(\xi;L)$ a rational function in $\xi$ with possible poles only at the poles of $a$,$b$ $c$ with degree at most $\text{max}(\text{deg}(a),\text{deg}(b),\text{deg}(c))-2$. 
Then, in the case where the Lax pair depends on some small parameter (in our case $\hbar$), they also prove that under the assumption that the correlation functions have a series expansion in this small parameter, then one can break \eqref{tyuu} into the usual standard loop equations arising in matrix models. In particular, the natural spectral curve arising is the one that we introduced earlier: the limit of the characteristic polynomial of $\mathcal{D}$ when the small parameter tends to $0$. One could think that since the correlation functions satisfy the loop equations, then applying the topological recursion to the limiting spectral curve should reconstruct every correlation function defined by \eqref{deter}. Unfortunately, this is not always the case because the loop equations may have several (and usually many) solutions. Thus one needs some sufficient criteria in order to ensure that the correlation functions obtained from the topological recursion of \cite{EO} applied to the limiting curve (which by definition are the one we construct on the matrix model side of the problem) will be the same as the one obtained from the determinantal formulas \eqref{deter}. In other words, are there some additional criteria to ensure that the loop equations have a unique solution? The paper \cite{BE} answers positively to the question and give us some sufficient criteria to ensure uniqueness:

\medskip
\underline{Hypothesis 1}:
If we denote by $\hbar$ the ``small" parameter, then the correlation functions $W_n$ defined by the determinantal formulas must have a series expansion at $\hbar=0$ of the form:
\beq W_n=\sum_{g=0}^\infty W_n^{(g)} \hbar^{n+2g-2}\eeq

\medskip
\underline{Hypothesis 2}:
Let $\underset{\hbar\to 0}{\lim}\det(Y-\mathcal{D})=0 \Leftrightarrow Y^2=E(t,\xi)$ be the limiting spectral curve. Then $\xi\to E(t,\xi)$ is algebraic and may have odd or even zeros. By construction, $W_n^{(g)}$ will possibly have branchcut singularities at odd zeros of $E$ and possibly pole singularities at even zeros of $E$. The second hypothesis consists in saying that $W_n^{(g)}$ have no pole at even zeros of $E$.

\medskip
\underline{Hypothesis 3}: We assume that the $W_n^{(g)}$ have fixed filling fractions, i.e.:
\beq \oint_{\mathcal{A}_i} W_n^{(g)}=0\eeq
for the homology cycles $\mathcal{A}_i$ defined on the Riemann surface given by the limiting spectral curve.

If the $3$ hypotheses are satisfied we get the following theorem (proved by \cite{BE}):
\begin{proposition} If the previous three hypotheses are satisfied then the functions $W_n^{(g)}$ constructed by the determinantal formulas \eqref{deter} are identical to the correlation functions constructed through the topological recursion applied to the limiting spectral curve $\lim_{\hbar\to 0}\det(y-\mathcal{D})=0$. The results also extend to $\ln \tau$ which corresponds to the case $n=0$.
\end{proposition}
 
Since we already know that the two spectral curves computed with each formalism are the same, we only need to prove the validity of the three hypotheses to prove that the correlations functions and $\tau$-functions computed on both sides are identical.

\subsection{Validation of hypothesis $3$}

Hypothesis $3$ is obvious. Indeed, since the limiting spectral curve is of genus $0$ (there is only one cut) then the only filling fraction is automatically fixed and therefore hypothesis $3$ is empty and trivially verified.

\subsection{Validation of hypothesis $2$}

Validation of hypothesis $2$ is more difficult. Indeed, we need to prove that the Lax pair wave functions have no singularity at $\xi=\frac{1}{2}$ which is not obvious. Up to the rescaling $t=-\frac{2is}{\pi}$, the problem can be directly studied through the Lax pair given earlier. The analysis is detailed in appendix \ref{AppendixB} where we prove the following:

\begin{theorem} \label{polestructure}
If we denote by:
\beq \psi(t,\xi)=\text{exp}\left(\sum_{k=-1}^\infty \psi_k(t,\xi) \hbar^{k}\right)=\text{exp}\left(\sum_{k=-1}^\infty \frac{\psi_k(\xi)}{t^k} \hbar^{k}\right)    \eeq
the WKB expansion of the wave functions, then we have $\psi_{-1}(t,\xi)=\pm y(t,\xi)$. Moreover, the only possible finite singularities involved in $\xi\mapsto\psi_k(t,\xi)$ are located at $\xi=0,1$. In particular they are not singular at $\xi=\frac{1}{2}$. These results also apply to the three other wave functions $\phi(t,\xi), \td{\psi}(t,\xi)$ and $\td{\phi}(t,\xi)$. 
\end{theorem}

With the help of the previous theorem, it is easy to see that the kernel $K(\xi_1,\xi_2)$ will have an asymptotic series expansion in $\hbar$ (the exponential part of the WKB expansion of the wave functions cancels as proved in appendix \ref{AppendixParity} and the rest of exponential can be expanded around $\hbar=0$) and that all orders will have no singularity at $\xi=\frac{1}{2}$. Thus, the determinantal formulas \eqref{deter} will lead to correlation functions $W_n$ whose $W_n^{(g)}$ asymptotic expansion in $\hbar$ have no singularities at $\xi=\frac{1}{2}$. Hence, as soon as theorem \eqref{polestructure} is proved, hypothesis $2$ is verified.  

\subsection{Validation of hypothesis $1$}

Hypothesis $1$ is subtle and had sometimes been overlooked in some papers. In fact, from the existence of an asymptotic expansion of the $\sigma(t)$ function, it is straightforward to see that the determinantal formulas will lead to a $\hbar$ asymptotic expansion for $W_n$. However two crucial points are still missing. The first one is the order of the leading order which is claimed to be $\hbar^{n-2}$ for $W_n$. Then the second one is the parity, i.e. that only even powers of $\hbar$ will be involved (we have in the end an expansion in $\hbar^2$ and not $\hbar$). 

\medskip

The justification of the parity in $\hbar$ is directly related to the following theorem:
\begin{theorem}\label{TheoremParity} Let us denote:
\bea
\psi(t,\xi)&=&\sum_{k=-1}^\infty \psi_k(t,\xi)\hbar^k=\sum_{k=-1}^\infty \frac{\psi_k(\xi)\hbar^k}{t^k}\cr
\phi(t,\xi)&=&\sum_{k=-1}^\infty \phi_k(t,\xi)\hbar^k=\sum_{k=-1}^\infty \frac{\phi_k(\xi)\hbar^k}{t^k}\cr
\td{\psi}(t,\xi)&=&\sum_{k=-1}^\infty \td{\psi}_k(t,\xi)\hbar^k=\sum_{k=-1}^\infty \frac{\td{\psi}_k(\xi)\hbar^k}{t^k}\cr
\td{\phi}(t,\xi)&=&\sum_{k=-1}^\infty \td{\phi}_k(t,\xi)\hbar^k=\sum_{k=-1}^\infty \frac{\td{\phi}_k(\xi)\hbar^k}{t^k}\cr
\eea
the asymptotic series of our wave functions. Then we have:
\beq \td{\phi}(t,\xi)=\psi(-t,\xi) \Leftrightarrow \forall\,  k\geq -1\,:\, \td{\phi}_k(\xi)=(-1)^k\psi_k(\xi)\eeq
and
\beq \td{\psi}(t,\xi)=\phi(-t,\xi) \Leftrightarrow \forall\,  k\geq -1\,:\, \td{\psi}_k(\xi)=(-1)^k\phi_k(\xi)\eeq
Moreover:
\beq \psi_{-1}(\xi)=\td{\psi}_{-1}(\xi)\, \text{ and } \phi_{-1}(\xi)=\td{\phi}_{-1}(\xi)\eeq
\end{theorem}
The proof of this theorem as well as the connection with the parity of $W_n$ are given in appendix \ref{AppendixParity}. The proof uses the invariance $t\leftrightarrow -t$ as well as symmetry/antisymmetry arguments when expanding the determinantal formulas.


Then, we also need to prove:
\bt\label{thWnleading}
The leading order of $W_n$ is:
\beq
W_n = O(\hbar^{n-2}).
\eeq
\et
The proof of this theorem is given in appendix \ref{AppendixthWnleading}. The proof is based on the introduction of a ``loop insertion operator" which acting on $W_n$ gives $W_{n+1}$:
\beq
\delta_{x_{n+1}} W_n(x_1,\dots,x_n) = W_{n+1}(x_1,\dots,x_n,x_{n+1})
\eeq
and we have to prove that $\delta_x$ adds a power of $\hbar$. This is proved using the fact that ${\cal R}(t,\xi)$ is linear in $\xi$.

\subsection{Conclusion and outlook}

With the validation of the three hypotheses, we conclude with the following theorem:
\begin{theorem} \label{MainTheo}
The correlation functions $W_n^{(g)}$ generated by the topological recursion \eqref{Toprec} are identical to the determinantal formulas coming from the Lax pair \eqref{deter}. Moreover, the symplectic invariants $F^{(g)}$ computed by the topological recursion (that generates the logarithm of the partition function of the matrix model) are generating the $\tau$ function of Painlevé $5$:
\beq \forall \,k\geq 0\,: \,\, \hat{\tau}_k(t)=F^{(k)}(s) \eeq
\end{theorem}

Hence, we recover here the standard bulk universality results directly from the topological recursion of \cite{EO} and determinantal formulas of \cite{BE}. In another words, we have just provided an alternative proof of the universality in matrix models in the bulk through the loop equations formalism. 

Also, we have proved that the topological recursion and geometric invariants introduced in \cite{EO} can contain all universal limit laws of random matrices. It had been known for edge universal laws, like Tracy-Widom and its Airy kernel in \cite{BE}, but so far this was not known for the bulk gap probability.

Eventually, this work opens the way to a generalization to higher dimensional Lax pair and critical points of type $(p,q)$ in matrix models that could be studied with the same method. Additionally, on the integrable system side, one could think about using the same approach on the other Painlevé equations. So far only Painlevé $2$ and $5$ have been treated in details but a complete study of the other $4$ equations may lead to interesting results. In a more general context, the link between Painlevé $5$ and the topological recursion applied to a simple spectral curve is typical of a mirror symmetry models. We believe that our approach could be used to prove that the asymptotics of Jones polynomials are given by the topological recursion of a simple curve (the associated A-polynomial). This problem is known to be difficult and even if one can compute numerically the first orders on both sides and observe that they match so far nobody has developed a way to prove this observation even for at leading order (volume conjecture). Moreover, it is known that the Hitchin systems underlying knot theory have a Lax pair formulation from Turaev-Reshetikin so that both sides of the problem are similar to the case presented in this article. In this spirit, having a proof for the case of Painlevé $5$ or any other integrable system is helpful and is a small step in this direction.

\section{Acknowledgements}

We would like to thank G. Borot, R. Conte, S. Ribault for valuable discussions on this topic.
The work realized by O. Marchal was performed within the framework of the LABEX MILYON (ANR-10-LABX-0070) of Université de Lyon, within the program ``Investissements d'Avenir" (ANR-11-IDEX-0007) operated by the French National Research Agency (ANR). The work of B.E. is partly supported by the Quebec government through the FQRNT fund, and by the ERC Knot-Fields. The authors thank the unknown referee for suggesting improvements and its patience regarding normalization factors.

\begin{appendices}

\section{\label{AppendixA} Recursion formula for the $\hbar$ expansion of the $\sigma$-function}

In this appendix, we shall prove the recursion \eqref{recursion}. We define:
\beq \sigma(t)=\sum_{k=0}^\infty \pi^{2k}\sigma_{k}t^{2-2k}\hbar^{2k}\eeq
Therefore we find:
\beq \ddot{\sigma}(t)^2=\sum_{k=0}^\infty \pi^{2k}\hbar^{2k}t^{-2k}\sum_{j=0}^k (2-2j)(1-2j)(2-2k+2j)(1-2k+2j)\sigma_{j}\sigma_{k-j}\eeq
i.e.:
\beq -\pi^2\hbar^2t^2\ddot{\sigma}(t)^2=-\sum_{k=1}^\infty \pi^{2k}\hbar^{2k}t^{4-2k}\sum_{j=0}^{k-1}(2-2j)(1-2j)(4-2k+2j)(3-2k+2j)\sigma_{j}\sigma_{k-j-1}\eeq
A straightforward computation also gives:
\beq 4(\dot{\sigma}(t))^2+\sigma(t)-t\dot{\sigma}(t)=\sum_{k=0}^\infty \hbar^{2k}\pi^{2k}t^{2-2k}\left((2k-1)\sigma_{k}+4\sum_{j=0}^k (2-2j)(1-k+j)\sigma_{j}\sigma_{k-j}\right)
\eeq
and 
\beq t\dot{\sigma}(t)-\sigma(t)=\sum_{k=0}^\infty \hbar^{2k}\pi^{2k}(1-2k)\sigma_{k}\eeq
Inserting back these series expansion in $\hbar$ into the differential equation
\beq -\pi^2\hbar^2t^2\ddot{\sigma}^2+(t\dot{\sigma}-\sigma)(t\dot{\sigma}-\sigma-4\dot{\sigma}^2)=0\eeq
leads to ($\forall k\geq 1$):
\bea &&\sum_{j=0}^{k-1}(2-2j)(1-2j)(4-2k+2j)(3-2k+2j)\sigma_{j}\sigma_{k-1-j}=\cr
&& \sum_{i=0}^k\left((2i-1)(1-2k+2i)\sigma_{i}\sigma_{k-i}+8\sum_{j=0}^i(1-j)(2-2i+2j)(1-2k+2i)\sigma_{j}\sigma_{i-j}\sigma_{k-i}\right)\cr
\eea
Isolating $\sigma_{k}$ then gives \eqref{recursion}.

\section{\label{AppendixB} Singularities of the wave functions: Proof of theorem \eqref{polestructure}}

Let us denote the wave functions by $\Psi(t,\xi)=\begin{pmatrix}
\psi(t,\xi)&\phi(t,\xi) \\
\td{\psi}(t,\xi)&\td{\phi}(t,\xi)
\end{pmatrix}$
In order to have compact notations for computations, we introduce the following definition (we dropped the factor $-\pi$ in front of $\hbar$ for simplification but of course it can be inserted back by redefining $\hbar\to -\pi \hbar$ in what follows):
\bea \mathcal{D}(t,\xi)&=&\begin{pmatrix}
\alpha(t,\xi)& \beta(t,\xi)\\
\gamma(t,\xi)& -\alpha(t,\xi)\\
\end{pmatrix}\cr
\mathcal{R}(t,\xi)&=&\begin{pmatrix}
\frac{\xi}{2}& \mu(t)\\
\nu(t)& -\frac{\xi}{2}\\
\end{pmatrix}\cr
\frac{\partial f}{\partial \xi} &\overset{\text{def}}{=}& f'(t,\xi)\cr
\frac{\partial f}{\partial t} &\overset{\text{def}}{=}& \dot{f}(t,\xi)\cr
\eea
According to \eqref{Lax}, the identification is given by
\bea \alpha(t,\xi)&=&\frac{t}{2}+\frac{z(t)}{\xi}-\frac{z(t)}{\xi-1}\cr
\beta(t,\xi)&=&-\frac{u(t)z(t)}{\xi}+\frac{u(t)y(t)z(t)}{\xi-1}\cr
\gamma(t,\xi)&=&\frac{z(t)}{u(t)\xi}-\frac{z(t)}{u(t)y(t)(\xi-1)}\cr
\mu(t)&=&-\frac{u(t)z(t)}{t}+\frac{u(t)y(t)z(t)}{t}\cr
\nu(t)&=&\frac{z(t)}{u(t)t}-\frac{z(t)}{u(t)y(t)t}\cr
\eea
Moreover from \eqref{sysyzu}, the asymptotic expansion of $\sigma(t)$ and the fact that $z=-\frac{d}{dt}\sigma(t)$, it is straightforward to see that the functions $y(t),u(t)$ and $z(t)$ will have an asymptotic expansion of the form:
\bea \label{you}z(t)&=&\sum_{k=0}^\infty\frac{z_k}{t^{2k-1}}\hbar^{2k} =-\frac{t}{8}+\frac{2\hbar^4}{t^3}+\frac{16\hbar^6}{t^5}+\frac{25152\hbar^8}{t^7}+O(\hbar^{10})\cr
y(t)&=&\sum_{k=0}^\infty \frac{y_k}{t^k}\hbar^k=-1-\frac{4i\hbar}{t}+\frac{8\hbar^2}{t^2}+\frac{32\hbar^4}{t^4}-\frac{320\hbar^5}{t^5}+O(\hbar^6)\cr
\frac{d}{dt}\log u(t)&=&\sum_{k=0}^\infty \frac{u_k}{t^{2k}}\hbar^{2k-1}=-\frac{i}{2\hbar}+\frac{2i\hbar}{t^2}+\frac{16i\hbar^3}{t^4}+O(\hbar^5)\cr
\eea

From there, it is easy to observe the following:
\begin{itemize} 
\item $\alpha(t,\xi)$ has an asymptotic expansion of the form $\alpha=\underset{k=0}{\overset{\infty}{\sum}} \alpha_k \hbar^k$
\item $\beta\gamma$ has an asymptotic expansion in $\hbar$ starting at $\hbar^0$  (because the $u(t)$ function cancels)
\item $\mu\nu$ has an asymptotic expansion in $\hbar$ starting at $\hbar^0$ (because the $u(t)$ function cancels)
\item Since $u(t)$ only depends on $t$ we have:
$$\frac{\beta'}{\beta}=\frac{(\xi-1)^2-y(t)\xi^2}{\xi(\xi-1)(1+\xi(y(t)-1))}=-\frac{2\xi^2-2\xi+1}{2\xi(\xi-1)}+O(\hbar)=\left(\frac{\beta'}{\beta}\right)_0+O(\hbar)$$
\item When we get a $t$ derivative, the situation is more complex:
\bea\frac{\dot{\mu}}{\mu}&=& \frac{u_0}{\hbar}+ O(\hbar)\cr
\frac{\dot{\nu}}{\nu}&=& -\frac{u_0}{\hbar}+ O(\hbar)\cr
\mu(t)\nu(t)=-\frac{1}{16}+O(\hbar^2)
\eea
\end{itemize}

Eventually, we will sometimes omit to write down explicitly the dependence regarding the variables $(t,\xi)$. Moreover, we will denote with an index $i$ the power of $\hbar$ involved in the series expansion around $\hbar=0$ (for example $\beta_i$ will be the term in $\hbar^i$ of the function $\beta(t,\xi)$).

\subsection{The detailed proof for $\psi(t,\xi)$}

With these notations, the characteristic equations of the Lax system are:
\beq \label{secxi}
\hbar^2\psi''=\hbar^2\frac{\beta'}{\beta} \psi' -\left(i\hbar\alpha'-i\hbar\alpha \frac{\beta'}{\beta}+\alpha^2+\beta\gamma\right)\psi\eeq
and
\beq \label{sect}
\hbar^2\ddot{\psi}=\hbar^2 \frac{\dot{\mu}}{\mu}\dot{\psi}-\left(\frac{\xi^2}{4}+\mu\nu-i\hbar\frac{\dot{\mu}\xi}{2\mu}\right)\psi
\eeq
From the general form of the characteristic equations, we can perform a WKB expansion in $\hbar$ order to determine $\psi$. Therefore we define:
\beq \label{opu}\psi(t,\xi)=\exp\left(\sum_{k=-1}^\infty \psi_k(t,\xi)\hbar^k\right)\eeq
where the functions $\psi_k$ are independent of $\hbar$. We can now determine the different functions involved in the WKB expansion of $\psi$.

\medskip
\underline{Order $\hbar^0$:}
Using \eqref{secxi}, we find at order $\hbar^0$:
\beq \label{tyu}\left(\psi_{-1}'(t,\xi)\right)^2=-(\alpha_0^2+(\beta\gamma)_0)=-\frac{t^2(2\xi-1)^2}{16\xi(\xi-1)}=y^2(t,\xi)\eeq
and
\beq \left(\dot{\psi_{-1}}(t,\xi)\right)^2+\frac{i}{2}\dot{\psi_{-1}}(t,\xi)=-\frac{\xi^2}{4}+\frac{1}{16}+\frac{\xi}{4}\eeq
This result was expected since $\alpha_0^2+(\beta\gamma)_0$ is the leading order of the characteristic polynomial of $\mathcal{D}$ which by definition is given as the square of the spectral curve. The two equations are compatible and give the following answer:
 
\beq\label{dotf} \encadremath{ \psi_{-1}(t,\xi)=-\frac{it}{4}-\frac{it}{2}\sqrt{\xi(\xi-1)} }\eeq
In a similar way (since $\phi(t,\xi)$ satisfies the same characteristic equation as $\psi(t,\xi)$ but with the sign of $\phi'_{-1}(t,\xi)$ changed):
\beq\label{dotff} \encadremath{ \phi_{-1}(t,\xi)=-\frac{it}{4}+\frac{it}{2}\sqrt{\xi(\xi-1)} }\eeq

\medskip
\underline{Order $\hbar^{1}$:}
The next step is to look at order $\hbar$ for both characteristic equations. The first one leads to:
\beq \label{eqq1} 2\psi_{0}'\psi_{-1}'=-\psi_{-1}''+\psi_{-1}'\left(\frac{\beta'}{\beta}\right)_0-i\alpha_0'+i\left(\alpha\frac{\beta'}{\beta}\right)_0+(\alpha^2+\beta\gamma)_1\eeq
The differential equation in $t$ becomes trivial and gives:
\beq \dot{\psi_0}=0\eeq
which indicates that $\psi_0$ is independent of $t$.
The $\xi$-differential equation can be explicitly solved and we find:
\beq \encadremath{\psi_0(\xi)=-\frac{1}{4}\ln\left(\xi(\xi-1)\right)+\frac{1}{2}\ln\left(\xi-\frac{1}{2}+\sqrt{\xi(\xi-1)}\right)}\eeq
In particular we see that the only $\xi$-dependent singularities are at $\xi=0,1$ but that $\xi=\frac{1}{2}$ is regular. This result is rather non-trivial from \eqref{eqq1}. Indeed, since $\psi_{-1}$ has a simple zero at $\xi=\frac{1}{2}$ we would expect $\psi_{0}'$  to have a simple pole there, thus giving a log singularity. But it can be checked that the r.h.s. of \eqref{eqq1} has a simple zero at $\xi=\frac{1}{2}$ thus canceling the possible singularity.

\medskip
\underline{Conclusion}: \textbf{The function $\xi \mapsto \psi_0(\xi)$ may only have branchcut singularities at $\xi=0,1$}. In particular, it does not have singularities at $\xi=\frac{1}{2}$.

\medskip
\medskip

\underline{Order $\hbar^2$}: The $t$ differential equation simplifies into:
\beq \dot{\psi}_1=-\frac{i}{4t^2\sqrt{\xi(\xi-1)}}\eeq
and the $\xi$ differential equation is:
\beq 2\psi_{-1}'\psi_1'=-\psi_0''-(\psi_0')^2+\left(\frac{\beta'}{\beta}\right)_0\psi_0'+\left(\frac{\beta'}{\beta}\right)_1\psi_{-1}'-i\alpha_1'+i\left(\alpha\frac{\beta'}{\beta}\right)_1-(\alpha^2+\beta\gamma)_2
\eeq
Solving both equations leads to:
\beq \encadremath{\psi_1(t,\xi)=\frac{i}{4t\sqrt{\xi(\xi-1)}} }\eeq

\medskip
\medskip

\underline{Order $\hbar^3$}: With the same method we find:
\beq \encadremath{\psi_2(t,\xi)=-\frac{(2\xi-1)(2\xi-1-2\sqrt{\xi(\xi-1)})}{4t^2\xi(\xi-1)} }\eeq
In particular we observe that it vanishes linearly at $\xi=\frac{1}{2}$

\medskip
\medskip

\underline{General order $\hbar^n$:}
The $\xi$-differential equation gives:
\beq \eqref{bla} 2\psi_{-1}'\psi_{n-1}'=-\psi_{n-2}''-\sum_{i=0}^{n-2}\psi_i'\psi_{n-2-i}'+\sum_{k=0}^{n-2}\left(\frac{\beta'}{\beta}\right)_k\psi_{n-2-k}'
-i\alpha'_{n-1}+i\left(\alpha\frac{\beta'}{\beta}\right)_{n-1}-(\alpha^2+\beta\gamma)_{n-2}\eeq
which by induction gives $\psi_{n-1}'$. We observe here that in order to get $\psi_{n-1}'$, we must divide by $\psi_{-1}'$ which unfortunately has a simple zero at $\xi=\frac{1}{2}$. At least from general considerations, we can exclude by induction all other singularities, but from the $\xi$-differential equation it is not clear why $\psi_{n-1}$ should not have any singularity at $\xi=\frac{1}{2}$. In order to remove this potential singularity, we can use the $t$ differential equation.

\medskip

For the $t$ differential equation. We get:

\beq \left(2\dot{\psi}_{-1}-\left(\frac{\dot{\mu}}{\mu}\right)_{-1}\right)\dot{\psi}_{n-1}=-\ddot{\psi}_{n-2}-\sum_{i=0}^{n-2}\dot{\psi_i}\dot{\psi}_{n-2-i}+\sum_{k=0}^{n-2}\left(\frac{\dot{\mu}}{\mu}\right)_{k}\dot{\psi}_{n-2-k}-(\mu\nu)_{n}-\frac{i\xi}{2}\left(\frac{\dot{\mu}}{\mu}\right)_{n-1}
\eeq 

Since $\left(\frac{\dot{\mu}}{\mu}\right)_{-1}=-\frac{i}{2}$ and from \eqref{dotf} $\dot{\psi}_{-1}=-\frac{i}{4}-\frac{i}{2}\sqrt{\xi(\xi-1)}$ we get:
\beq \label{Generalt} -i\sqrt{\xi(\xi-1)}\dot{\psi}_{n-1}=-\ddot{\psi}_{n-2}-\sum_{i=0}^{n-2}\dot{\psi_i}\dot{\psi}_{n-2-i}+\sum_{k=0}^{n-2}\left(\frac{\dot{\mu}}{\mu}\right)_{k}\dot{\psi}_{n-2-k}-(\mu\nu)_{n}-\frac{i\xi}{2}\left(\frac{\dot{\mu}}{\mu}\right)_{n-1}
\eeq 
Therefore if we assume by induction that the $\psi_k(t,\xi)$ for $k\leq n-2$ do not have any singularity at $\xi=\frac{1}{2}$, we see that since $\dot{\psi}_{n-1}$ is regular at $\xi=\frac{1}{2}$ then the potential simple pole of ${\psi}_{n-1}'$ must have a constant residue ($t$-independent). However, it is not enough to completely remove constant ($t$-independent) simple poles at $\xi=\frac{1}{2}$ for which the $t$ differential equation does not bring anything. 

\medskip
\medskip

\underline{Using the explicit $t$-dependence}:  In section \eqref{tdependancee} we have seen how to rewrite the $\hbar$ series expansion in a large $t$ series expansion. In particular the general WKB form of \eqref{opu} shows that we should get the explicit $t$-dependence of the form:
\beq \label{tdependancee}\encadremath{\psi_k(t,\xi)=\frac{\psi_k(\xi)}{t^k} }\eeq
This is of course compatible with our results for the first orders presented earlier in this appendix. Moreover it is obvious that a solution like \eqref{tdependance} is correct in \eqref{Generalt} since we have:
\beq \encadremath{(\mu\nu)_n(t)=\frac{C_n}{t^n} \,\,\text{ and }  \left(\frac{\dot{\mu}}{\mu}\right)_{n}(t)=\frac{D_n}{t^{n+1}} }\eeq
where $C_n$ and $D_n$ are numbers.
Indeed, the first identity comes from:
\beq \mu(t)\nu(t)=\frac{z(t)}{t^2}\left(-2+y(t)+\frac{1}{y(t)}\right)=i\hbar\frac{z(t)}{t}\frac{d \log u(t)}{dt}=\frac{i\hbar}{t}\left(\sum_{k=0}^\infty z_k\frac{\hbar^{2k}}{t^{2k-1}}\right)\left(\sum_{k=0}^\infty u_k\frac{\hbar^{2k-1}}{t^{2k}}\right)\eeq
where we have used the expansion \eqref{you} for $\log u(t)$ and $z(t)$.
The second identity also comes from \eqref{you} and the fact that:
\bea \frac{\dot{\mu}}{\mu}(t)&=&\frac{d}{dt}\log u(t)-\frac{1}{t}+\frac{\dot{z}}{z}+\frac{\dot{y}}{y-1}\cr
&=&\sum_{k=0}^\infty \frac{u_k\hbar^{2k-1}}{t^{2k}}-\frac{1}{tz_0}\frac{\underset{k=1}{\overset{\infty}{\sum}} \frac{(2k-1)z_k\hbar^{2k}}{t^{2k}}}{1+\underset{k=2}{\overset{\infty}{\sum}} \frac{z_k\hbar^{2k}}{z_0t^{2k}}}- \frac{1}{(y_0-1)t}\frac{\underset{k=0}{\overset{\infty}{\sum}} \frac{ky_k\hbar^k}{t^k}}{1+\underset{k=1}{\overset{\infty}{\sum}} \frac{y_k\hbar^k}{(y_0-1)t^k}} 
\eea

\underline{Excluding the singularity at $\xi=\frac{1}{2}$}:
From the knowledge of the $t$-dependence of the $\psi_k(t,\xi)$ functions, we deduce from \eqref{Generalt}
\bea \label{Generalt2} i(n-1)\sqrt{\xi(\xi-1)}\psi_{n-1}(\xi)&=&-(n-2)(n-1)\psi_{n-2}(\xi)-\sum_{i=0}^{n-2}i(n-2-i)\psi_i(\xi)\psi_{n-2-i}(\xi)\cr
&&+\sum_{k=0}^{n-2}(n-2-k)D_k\psi_{n-2-k}(\xi)-C_n-\frac{i\xi D_{n-1}}{2}
\eea
which directly gives $\psi_{n-1}(\xi)$ and not only its time derivative. Since we only have to divide by $i(n-1)\sqrt{\xi(\xi-1)}$ which is regular at $\xi=\frac{1}{2}$, we get by induction that $\psi_{n-1}(\xi)$ do not have any singularity at $\xi=\frac{1}{2}$. In fact from \eqref{Generalt2} it is obvious to show by induction that $\psi(\xi)$ will only have singularities at $\xi\in\{0,1\}$. This ends the proof of theorem \eqref{polestructure} for $\psi(t,\xi)$.

\subsection{Adaptation of the proof for the other wave functions}

The previous proof can directly be adapted for the other wave functions $\td{\psi}$, $\phi$ and $\td{\phi}$. Indeed, the characteristic equations \eqref{secxi} and \eqref{sect} are the same for $\phi(t,\xi)$ and the proof is identical to the previous one. The only change for $\td{\psi}(t,\xi)$ is that we must choose the other sign of the spectral curve: $\td{\psi}_{-1}(t,\xi)=-\psi_{-1}(t,\xi)$ and $\td{\phi}_{-1}(t,\xi)=-\phi_{-1}(t,\xi)$. The rest of the proof remains identical.

\section{Proof of theorem \eqref{TheoremParity}\label{AppendixParity}}
We want to prove \eqref{TheoremParity}. In order to do that, we will use the invariance under the transformation $t\mapsto -t$. We will only detail here the relation between $\psi(-t,\xi)$ and $\td{\phi}(t,\xi)$ since the second one between $\td{\psi}(t,\xi)$ and $\phi(-t,\xi)$ can be immediately adapted from this one. Let's observe first that $\psi(t,\xi)$ and $\td{\phi}(t,\xi)$ are solutions of the characteristic equations:
\bea 
\hbar^2\psi''&=&\hbar^2\frac{\beta'}{\beta} \psi' -\left(i\hbar\alpha'-i\hbar\alpha \frac{\beta'}{\beta}+\alpha^2+\beta\gamma\right)\psi\cr
\hbar^2\ddot{\psi}&=&\hbar^2 \frac{\dot{\mu}}{\mu}\dot{\psi}-\left(\frac{\xi^2}{4}+\mu\nu-i\hbar\frac{\dot{\mu}\xi}{2\mu}\right)\psi\cr
\eea
and 
\bea \label{secuu}
\hbar^2\td{\phi}''&=&\hbar^2\frac{\gamma'}{\gamma} \td{\phi}' -\left(-i\hbar\alpha'+i\hbar\alpha \frac{\gamma'}{\gamma}+\alpha^2+\beta\gamma\right)\td{\phi}\cr
\hbar^2\ddot{\td{\phi}}&=&\hbar^2 \frac{\dot{\nu}}{\nu}\dot{\td{\phi}}-\left(\frac{\xi^2}{4}+\mu\nu+i\hbar\frac{\dot{\nu}\xi}{2\nu}\right)\td{\phi}\cr
\eea

From the invariance of the Chazy equation when $t\mapsto -t$ and the various connections between all the functions we get:
\bea \sigma(-t)&=&\sigma(t)\cr
z(-t)&=&-z(t)\cr
\log u(-t)&=&-\log u(t)\cr
y(-t)&=&\frac{1}{y(t)}\cr
\alpha(-t,\xi)&=&-\alpha(t,\xi)\cr
(\beta\gamma)(-t,\xi)&=&(\beta\gamma)(t,\xi)\cr
(\mu\nu)(-t)&=&(\mu\nu)(t)\cr
\frac{\dot{\mu}}{\mu}(-t)&=&\frac{\dot{\nu}}{\nu}(t)\cr
\frac{\beta'}{\beta}(-t)&=&\frac{\gamma'}{\gamma}(t)\cr
\eea

In order to avoid confusion with the derivatives, we will note in this section $G(t,\xi)=\psi(-t,\xi)$. Therefore, we get that \textbf{$G(t,\xi)$ satisfy the same characteristic equations as $\td{\phi}(t,\xi)$}. From the definition and result \eqref{dotf} we have $G_{-1}(t,\xi)=\psi_{-1}(-t,\xi)=\frac{it}{4}+\frac{it}{2}\sqrt{\xi(\xi-1)}$
On the other side, the leading order projection of the characteristic equations for $\td{\phi}$ leads to: 
\bea 
 \left(\td{\phi}_{-1}'(t,\xi)\right)^2=y^2(t,\xi)\cr
 \left(\dot{\td{\phi}}_{-1}(t,\xi)\right)^2-\frac{i}{2}\dot{\td{\phi}}_{-1}(t,\xi)&=&-\frac{\xi^2}{4}+\frac{1}{16}+\frac{\xi}{4}
\eea
By definition of the matrix $\Psi(t,\xi)$, we must choose $\td{\phi}_{-1}'(t,\xi)=-\psi_{-1}'(t,\xi)=\frac{it(2\xi-1)}{4\xi(\xi-1)}$. Hence the resolution of the two differential equations leads to:
\beq \td{\phi}_{-1}(t,\xi)=\frac{1}{2}it\sqrt{\xi(\xi-1)}+\frac{it}{4}=G_{-1}(t,\xi)\eeq
Therefore \textbf{the leading orders of $\td{\phi}$ and $G$ are the same}. The rest of the proof follows by induction using recursion formulas similar to \eqref{Generalt} and \eqref{bla} adapted to our characteristic equations \eqref{secuu} that are satisfied by both $\td{\phi}$ and $G$. One can also use the uniqueness of the solution of second order linear differential equations like \eqref{secuu} with prescribed leading order. Anyway by induction it proves that:
\beq \td{\phi}(t,\xi)=\psi(-t,\xi) \,\Leftrightarrow \forall\, k\geq-1\,:\, \td{\phi}_k=(-1)^k \psi_k\eeq
Using similar arguments we also get:
\beq \td{\psi}(t,\xi)=\phi(-t,\xi) \,\Leftrightarrow \forall\, k\geq-1\,:\, \td{\psi}_k=(-1)^k \phi_k\eeq
This proves \ref{TheoremParity}.

\medskip
\underline{Justification of the parity of the series expansion of $W_n$}:

We explain here how we can use theorem \ref{TheoremParity} in order to justify the form of the asymptotic series of $W_n$ which is claimed to be:
\beq W_n(\xi_1,\dots,\xi_n)=\sum_{g=0}^\infty W_n^{(g)}(\xi_1,\dots,\xi_n)\hbar^{n-2+2g}\eeq
First note that we have (see \eqref{dotf} and \eqref{dotff} and \ref{TheoremParity}):
\bea \label{LeadingOrders}
\psi_{-1}(t,\xi)&=&-\frac{it}{2}\sqrt{\xi(\xi-1)}-\frac{it}{4}\cr
\phi_{-1}(t,\xi)&=&\frac{it}{2}\sqrt{\xi(\xi-1)}-\frac{it}{4}=-\psi_{-1}(t,\xi)-\frac{it}{2}\cr
\td{\psi}_{-1}(t,\xi)&=&-\frac{it}{2}\sqrt{\xi(\xi-1)}+\frac{it}{4}=\psi_{-1}(t,\xi)+\frac{it}{2}\cr
\td{\phi}_{-1}(t,\xi)&=&\frac{it}{2}\sqrt{\xi(\xi-1)}+\frac{it}{4}=-\psi_{-1}(t,\xi)\cr
\eea
Additionally, one can also check that:
\bea \label{SecondOrders}
\psi_0(\xi)&=&\frac{1}{2}\ln \left(\xi-\frac{1}{2}+\sqrt{\xi(\xi-1)}\right)-\frac{1}{4}\ln\left(\xi(\xi-1)\right)+\frac{1}{2}\ln 2+\frac{i\pi}{2}\cr
\phi_0(\xi)&=&\frac{1}{2}\ln \left(\xi-\frac{1}{2}-\sqrt{\xi(\xi-1)}\right)-\frac{1}{4}\ln\left(\xi(\xi-1)\right)+\frac{1}{2}\ln 2+i\pi\cr
\td{\psi}_0(\xi)&=&\phi_0(\xi)\cr
\td{\phi}_0(\xi)&=&\psi_0(\xi)
\eea
The constants here are chosen for convenience in order to fix $\det \Psi=1$. Moreover, we remind the reader that according to \cite{BE}, the determinantal formulas are given by \eqref{deter}. First we observe that the result for $W_1(\xi)=\psi'(t,\xi)\td{\phi}(t,\xi)-\td{\psi}(t,\xi)\phi'(t,\xi)$ is obvious from \eqref{tyu}. Moreover, a straightforward computation (using the $\det \Psi(t,\xi)=1$) gives:
\bea W_2(\xi_1,\xi_2)&=&-\frac{1}{(\xi_1-\xi_2)^2}\left(\psi(t,\xi_1)\td{\psi}(t,\xi_2)-\td{\psi}(t,\xi_1)\psi(t,\xi_2)\right)\left(\phi(t,\xi_1)\td{\phi}(t,\xi_2)-\td{\phi}(t,\xi_1)\phi(t,\xi_2)\right)\cr
&=&\frac{1}{(\xi_1-\xi_2)^2}\left(\psi(t,\xi_1)\td{\psi}(t,\xi_2)-\td{\psi}(t,\xi_1)\psi(t,\xi_2)\right)\cr
&&\left(\psi(-t,\xi_1)\td{\psi}(-t,\xi_2)-\td{\psi}(-t,\xi_1)\psi(-t,\xi_2)\right)\cr
\eea
Since we get here a symmetric expression in $t$ then the series expansion may only involve even powers of $t$ and as a consequence only even powers of $\hbar$.

Let's now deal with the general case $n>2$:
\beq \label{determinn} W_n(\xi_1,\dots,\xi_n)=(-1)^{n+1}\sum_{\tau \text{ cycles}}\prod_{i=1}^nK(\xi_i,\xi_{\tau(i)})\eeq
which gives for example:
\beq W_3(\xi_1,\xi_2,\xi_3)=K(\xi_1,\xi_2)K(\xi_2,\xi_3)K(\xi_3,\xi_1)+ K(\xi_1,\xi_3)K(\xi_3,\xi_2)K(\xi_2,\xi_1)\eeq
We rewrite the WKB expansion of the $\psi(t,\xi)$ functions as (using \eqref{LeadingOrders}):
\bea 
\psi(t,\xi)=e^{\frac{\psi_{-1}(\xi)t}{\hbar}}\left(\sum_{k=0}^\infty \frac{\Psi_k(\xi)\hbar^k}{t^k}\right)\cr
\phi(t,\xi)=e^{\frac{-\psi_{-1}(\xi)t-\frac{it}{2}}{\hbar}}\left(\sum_{k=0}^\infty \frac{\Phi_k(\xi)\hbar^k}{t^k}\right)\cr
\td{\psi}(t,\xi)=e^{\frac{\psi_{-1}(\xi)t+\frac{it}{2}}{\hbar}}\left(\sum_{k=0}^\infty \frac{(-1)^k\Phi_k(\xi)\hbar^k}{t^k}\right)\cr
\td{\phi}(t,\xi)=e^{\frac{-\psi_{-1}(\xi)t}{\hbar}}\left(\sum_{k=0}^\infty \frac{(-1)^k\Psi_k(\xi)\hbar^k}{t^k}\right)\cr
\eea
This rewriting is merely the expansion of the regular part of the former WKB expansion. We find it more convenient for the following proof to use such expressions. The functions $\Psi_k(\xi)$, $\Phi_k(\xi)$, $\td{\Psi}_k(\xi)$ and $\td{\Phi}_k(\xi)$ can be directly computed from the former $\psi_k(\xi)$, $\phi_k(\xi)$, $\td{\psi}_k(\xi)$ and $\td{\phi}_k(\xi)$ functions. Then we get:
\beq \label{KKK} K(\xi_1,\xi_2)=e^{\frac{it\left(\sqrt{\xi_2(\xi_2-1)}-\sqrt{\xi_1(\xi_1-1)}\right) }{2\hbar}}\left(\sum_{k=0}^\infty \frac{K_k(\xi_1,\xi_2)\hbar^k}{t^k}\right)\eeq
with
\beq K_k(\xi_1,\xi_2)=\frac{1}{\xi_1-\xi_2}\left(\sum_{i+j=k}(-1)^j\Psi_i(\xi_1)\Psi_j(\xi_2)-\sum_{i+j=k}(-1)^j\Phi_i(\xi_1)\Phi_j(\xi_2)\right)\eeq
In particular we observe the following symmetry: \textbf{$K_k(\xi_1,\xi_2)$ is a symmetric function of $(\xi_1,\xi_2)$ when $k$ is odd and is an antisymmetric function of $(\xi_1,\xi_2)$ when $k$ is even.}
This observation leads to the following ones:
\begin{itemize}
\item For each term of the sum over cycles in \eqref{determinn}, the product of exponential terms involved cancel. Indeed, each cycle uniquely involves each index in the first and second variable of one $K$ function and thus in the end they cancel. In particular this proves that each term in the sum over cycles has an asymptotic series in $\hbar$ even if $K(\xi_1,\xi_2)$ does not.
\item Let's now look at a given $W_n$. Since in the sum we deal with a cyclic product of $n$ functions $K(\xi_i,\xi_{\tau(i)})$  we get:
\beq \prod_{i=1}^nK(\xi_i,\xi_{\tau(i)})=\sum_{k=0}^\infty G_{k,\tau}(\xi_1,\dots,\xi_n)\frac{\hbar^k}{t^k}\eeq
with the following symmetry property: \textbf{If $n$ is odd then $G_{k,\tau}$ is an antisymmetric function of $(\xi_1,\dots,\xi_n)$ when $k$ is even while it is a symmetric function of $(\xi_1,\dots,\xi_n)$ when $k$ is odd. On the contrary, when $n$ is even then $G_{k,\tau}$ is an antisymmetric function of $(\xi_1,\dots,\xi_n)$ when $k$ is odd while it is a symmetric function of $(\xi_1,\dots,\xi_n)$ when $k$ is even.}
Indeed, for a cycle $\tau$, the symmetry of the product only depends on how many functions of each type (symmetric or antisymmetric) we multiply together.  
\end{itemize}
From the last point, we see that when we sum over all cycles then all antisymmetric functions $G_{k,\tau}(\xi_1,\dots,\xi_n)$ will add up to zero by symmetry. Therefore we get the expected result:
\medskip
\textbf{When $n$ is odd, then the series expansion of $W_n$ exists and only involves odd powers of $\hbar$ whereas when $n$ is even, then the series expansion of $W_n$ exists and only involves even powers of $\hbar$}

We stress here that it is only the entire sum over cycles that exhibits a series expansion in $\hbar^2$ but that each term taken separately only has a series expansion in $\hbar$. However our reasoning does not explain so far why the leading order of the expansion should start at $\hbar^{n-2}$. From what we said here, we only proved that it starts at $\hbar^0$ or $\hbar^1$ depending on the parity of $n$.

\section{Proof of theorem \ref{thWnleading}}\label{AppendixthWnleading}

So far, we have considered the Painlevé 5 Lax pair, with matrices ${\cal D}(t,\xi)$ and ${\cal R}(t,\xi)$ having simple poles at $\xi=0,1,\infty$ with given (vanishing) monodromies, and satisfying the Lax equation.
The most general solution can be expressed, modulo a gauge transformation, as \eqref{defDtxi} and \eqref{defRtxi}.

Here, we are going to introduce an ``insertion operator", which does not preserve the normalization of \eqref{defDtxi} and \eqref{defRtxi},
$\det\Psi(t,\xi)=1$ condition,
so we need to enlarge slightly our framework, and we only assume that:
\beq
{\cal R}(t,\xi) = \frac{\xi}{2}\,\sigma_3 + {\cal R}_0(t)
\quad , \quad
{\cal R}_0(t)=\frac{1}{2}\,\begin{pmatrix}
\rho(t) & \mu(t) \cr \nu(t) & -\rho(t)
\end{pmatrix}.
\eeq
Similarly
\beq
{\cal D}(t,\xi) = \frac{t}{2}\,\sigma_3 + \frac{{\cal A}_0(t)}{\xi}+\frac{{\cal A}_1(t)}{\xi-1}
\eeq

Following \cite{BE} we define the matrix:
\beq M(t,\xi)=
\Psi(t,\xi) \begin{pmatrix}
1 & 0 \cr 0 & 0
\end{pmatrix}
\Psi^{-1}(t,\xi)
\eeq
It is a rank $1$ projector:
\beq
M(t,\xi)^2=M(t,\xi)
\quad , \quad
\Tr M(t,\xi)=1
\quad , \quad
\det M(t,\xi)=0.
\eeq
We shall write it:
\beq
M(t,\xi)
=\frac{1}{2}\,{\rm Id} + \frac{1}{2}\,\begin{pmatrix}
h(t,\xi) & f(t,\xi) \cr
g(t,\xi) & -h(t,\xi)
\end{pmatrix}
\eeq
where $\det M(t,\xi)=0$ imposes
\beq
h^2(t,\xi) = 1 - f(t,\xi)g(t,\xi).
\eeq

It satisfies:
\beq
\hbar\,\partial_t\,M(t,\xi) = [{\cal R}(t,\xi),M(t,\xi)]
\eeq
i.e.
\bea\label{eqhbardRdt}
\hbar\,\dot f(t,\xi) & =& (x+\rho(t))f(t,\xi) - \mu(t) h(t,\xi)  \cr
\hbar\,\dot g(t,\xi) & =& -(x+\rho(t))g(t,\xi) + \nu(t) h(t,\xi)  \cr
\hbar\,\dot h(t,\xi) & =& \frac{1}{2}\,(\mu(t)g(t,\xi)-\nu(t)f(t,\xi) ) 
\eea

It is easy to see that it has an $\hbar$ expansion:
\beq
M(t,\xi) = \sum_k \frac{\hbar^k}{t^k}\, M_k(\xi).
\eeq

Then, observe that a rewriting of the determinantal formulas can be done using only the $M(t,\xi)$ matrix for $n\geq 2$ (theorem $2.1$ of \cite{BE}):
\beq\label{WnintermsofM} W_n(\xi_1,\dots,\xi_n)=(-1)^{n+1}\Tr \sum_{\tau \text{ cyclic}}\prod_{i=1}^n \frac{M(t,\xi_{\tau(i)})}{\xi_{\tau(i)}-\xi_{\tau(i+1)}}\eeq

\subsection{The insertion operator}

Let us consider the Picard-Vessiot differential ring $\mathbb B$ generated by the entries of $\Psi(t,\xi)$ as well as their $t$ derivatives, over the field of  rational functions $\mathbb C(\xi)$ (constant in $t$).
Also consider the $n$-variable analog $\mathbb B_n$ of $\mathbb B$ generated by the entries of $\Psi(t,\xi_1),\dots,\Psi(t,\xi_n)$  as well as their $t$ derivatives. And let $\mathbb B_\infty=\lim_{n\to\infty} \mathbb B_n$ its projective limit.

We define the following derivation acting in $\mathbb B_\infty$, in fact sending  $\mathbb B_n$ into $\mathbb B_{n+1}$.

\bd
$\delta_{\xi_{n+1}}:\mathbb B_n\to\mathbb B_{n+1}$ is a derivation (i.e. it satisfies Leibniz rule), defined by its action on generators:
\beq
\delta_{\xi_{n+1}} \Psi(t,\xi_i) = \frac{M(t,\xi_{n+1})}{\xi_i-\xi_{n+1}}\,\Psi(t,\xi_i) + \frac{f(t,\xi_{n+1})}{2\mu(t)}\sigma_3\,\Psi(t,\xi_i)
\eeq
where we recall that $f(t,\xi)=2\,M_{1,2}(t,\xi)$.

\ed

In order for this definition to make sense, i.e. for $\delta_\xi$ to be well defined, we need to check that $[\delta_{\xi_1},\delta_{\xi_2}]=0$, and that $[\delta_\xi,\partial_t]=0$.

For that purpose we first assume that $\delta$ is well defined and we deduce:

\begin{lemma}
\label{L0a} If $\delta$ is an insertion operator, we have
\bea
\delta_{\eta}K(\xi_1,\xi_2) & = & -K(\xi_1,\eta)K(\eta,\xi_2),  \\
\label{Pder}\delta_{\eta} M(\xi) & = & \left[\frac{M(\eta)}{\xi-\eta} + \frac{f(\eta)}{2\mu(t)}\,\sigma_3 ,M(\xi)\right],  \\
\label{Lder}\delta_{\eta} {\cal D}(\xi) & = & \left[\frac{M(\eta)}{\xi-\eta} + \frac{f(\eta)}{2\mu(t)}\,\sigma_3 ,{\cal D}(\xi)\right] - \frac{M(\eta)}{(\xi - \eta)^2}, \\
\label{detPsider}\delta_{\eta} \ln\det \mathbf{\Psi}(\xi) & = &   \frac{1}{\xi - \eta},   \\
\label{djq} \delta_{\eta} W_{n}(\xi_1,\dots,\xi_n) & = & {W}_{n + 1}(\eta,\xi_1,\dots,\xi_n).
\eea
\end{lemma}
This last property justifies the name ``insertion operator": it sends ${W}_n$ to ${W}_{n+1}$.

\medskip

Using this lemma, and in particular \eq{Pder} we have
\beq
\delta_\eta f(t,\xi) = 2\,\frac{[M(t,\eta),M(t,\xi)]_{1,2}}{\xi-\eta} + \frac{f(t,\eta)\,f(t,\xi)}{\mu(t)}\,
\eeq

The condition $[\delta_\xi,\partial_t]=0$ amounts to
\beq
\delta_\eta {\cal R}(t,\xi)=\hbar\,\partial_t \left(\frac{f(t,\eta)}{\mu(t)}\right)\, \frac{\sigma_3}{2}   + \frac{f(t,\eta)}{\mu(t)}\,\left[\frac{\sigma_3}{2},{\cal R}(t,\xi)\right]+ \left[M(t,\eta),\frac{{\cal R}(t,\xi)-{\cal R}(t,\eta)}{\xi-\eta}\right]
\eeq
i.e. using ${\cal R}(t,\xi)=\xi\frac{\sigma_3}{2}+ {\cal R}_0(t)$, this defines $\delta_\eta {\cal R}_0(t)$ as:
\beq
\delta_\eta {\cal R}_0(t) = \hbar\,\partial_t \left(\frac{f(t,\eta)}{\mu(t)}\right)\, \frac{\sigma_3}{2}   + \left[ \frac{\sigma_3}{2}, \frac{f(t,\eta)}{\mu(t)}\, {\cal R}_0(t) - M(t,\eta)\right]
\eeq
In particular
\beq
\delta_\eta\mu(t)  = 0
\quad , \quad
\delta_\eta\nu(t)  = g(t,\eta)- f(t,\eta)\,\frac{\nu(t)}{\mu(t)}.
\quad , \quad
\delta_\eta \rho(t) = \hbar\,\partial_t \left(\frac{f(t,\eta)}{\mu(t)}\right)
\eeq

we see that the condition $[\delta_\xi,\delta_\eta]=0$ amounts to verifying that:
\beq
\left(\delta_\xi \left(\frac{f(t,\eta)}{\mu(t)}\right) - \delta_\eta \left(\frac{f(t,\xi)}{\mu(t)}\right)\right)\,\frac{\sigma_3}{2}
- \left[ \frac{f(t,\xi)}{\mu(t)}\,\frac{\sigma_3}{2}, \frac{f(t,\eta)}{\mu(t)}\,\frac{\sigma_3}{2}\right]  = 0
\eeq
and one can easily check that this holds true.
Therefore the insertion operator $\delta$ is well defined.

\subsection{Restriction to the sub ring generated by $M$ and ${\cal R}$}

Let $\hat{\mathbb B}_n$ be the subring of $\mathbb B_n$ generated by the entries of $f(t,\xi_i)$, $g(t,\xi_i)$, $h(t,\xi_i)$, $i=1,\dots,n$, and $\mu(t)$. And let $\hat{\mathbb B}_\infty=\lim_{n\to\infty} \hat{\mathbb B}_n$ its projective limit.

We have seen in the previous paragraph, that the insertion operator sends $\hat{\mathbb B}_n$ to $\hat{\mathbb B}_{n+1}$.
It satisfies:
\begin{lemma}
The operator $\delta:\hat{\mathbb B}_n\to\hat{\mathbb B}_{n+1}$ is well defined (in other words the algebra generated by $\mu,f,g,h$ and $\delta$ closes). When acting on generators of $\hat{\mathbb B}_n$  it gives:
\label{LemmadeltaM}
\bea
\delta_\eta \mu(t) &=& 0 \cr
\delta_\eta f(t,\xi) &=& \frac{f(t,\xi)\,f(t,\eta)}{\mu(t)}- \frac{1}{\,\mu(t)^2}\,\frac{h(t,\xi)f(t,\eta)-h(t,\eta)f(t,\xi)}{\xi-\eta} \cr
\delta_\eta g(t,\xi) &=& -\,\frac{g(t,\xi)\,f(t,\eta)}{\mu(t)}+ \frac{1}{\,\mu(t)^2}\,\frac{h(t,\xi)g(t,\eta)-h(t,\eta)g(t,\xi)}{\xi-\eta} \cr
\delta_\eta h(t,\xi) &=&  \frac{1}{2\,\mu(t)^2}\,\frac{f(t,\xi)g(t,\eta)-f(t,\eta)g(t,\xi)}{\xi-\eta} 
\eea
\end{lemma}

Our goal now, will be to prove that the right hand sides, are in fact $O(\hbar)$.

\subsection{Proving the $\hbar^{n-2}$ property}

\smallskip

Define the following matrix $C(t,\xi)\in \hat{\mathbb B}_\infty$
\beq
C(t,\xi) = \frac{1 }{2\mu(t)}\,\frac{d}{dt}\,\begin{pmatrix}
f(t,\xi) & 0 \cr
-2 h(t,\xi) & -f(t,\xi)
\end{pmatrix}
\eeq
One can compute, using \eq{eqhbardRdt}:
\beq
2\mu(t) \hbar\,\ C(t,\xi) = \mu(t)\,{\rm Id} - 2\mu(t)\,M(t,\xi) + 2 f(t,\xi)\,{\cal R}(t,\xi)
\eeq
i.e.
\beq
M(t,\xi) = \frac{1}{2}\,{\rm Id} + \frac{f(t,\xi)}{\mu(t)} \, {\cal R}(t,\xi) - \hbar\,C(t,\xi)
\eeq

Now let us compute
\bea
\delta_\eta M(t,\xi) 
&=& \left[ \frac{M(t,\eta)}{\xi-\eta} + \frac{f(t,\eta)}{\mu(t)}\,\frac{\sigma_3}{2},M(t,\xi)\right] \cr
&=& \left[   \frac{f(t,\eta)}{\mu(t)}\,({\cal R}(t,\eta)+(\xi-\eta)\frac{\sigma_3}{2}),M(t,\xi)\right] - \hbar\,\left[\frac{C(t,\eta)}{\xi-\eta},M(t,\xi)\right] \cr
&=& \left[   \frac{f(t,\eta)}{\mu(t)}\,{\cal R}(t,\xi),M(t,\xi)\right] - \hbar\,\left[\frac{C(t,\eta)}{\xi-\eta},M(t,\xi)\right] \cr
&=& -\hbar\,\left[   \frac{f(t,\eta)}{\mu(t)}\,{\cal R}(t,\xi),C(t,\xi)\right] - \hbar\,\left[\frac{C(t,\eta)}{\xi-\eta},M(t,\xi)\right] \cr
\eea

This shows that $\delta_\eta$ acting in $\hat{\mathbb B}_\infty$, is proportional to $\hbar$.

\medskip

Hence from the last line of \eqref{WnintermsofM} we see that adding a new variable in $W_n$ for $n\geq 2$ lowers its leading order by at least one power of $\hbar$. Since we know that $W_2(\xi_1,\xi_2)$ is of order $\hbar^0$, we then conclude that the leading order of $W_n$ is at least of order $\hbar^{n-2}$ thus proving the corresponding part of Hypothesis $1$.

\end{appendices}

\end{document}